\documentclass[aps,pre,preprint,nofootinbib]{revtex4-2}

\usepackage{graphicx,amssymb}% Include figure files
\usepackage{dcolumn}% Align table columns on decimal point
\usepackage{amsmath,amsfonts}% popular packages from the American Mathematical Society
\usepackage{bm}% Bold Math package
\usepackage{fancyhdr,calc}
\usepackage{endnotes}
\usepackage[normalem]{ulem}
\usepackage{color}
\usepackage{xcolor}
\usepackage{multirow} % to tables
\newcommand{\flc}[1]{\textcolor{blue}{#1}}

\newcommand{\beq}{\begin{equation}}
\newcommand{\eeq}{\end{equation}}
\newcommand{\bea}{\begin{eqnarray}}
\newcommand{\eea}{\end{eqnarray}}

\newcommand{\beqn}{\begin{equation*}}
\newcommand{\eeqn}{\end{equation*}}
\newcommand{\bean}{\begin{eqnarray*}}
\newcommand{\eean}{\end{eqnarray*}}
\newcommand{\bal}{\begin{align}}
\newcommand{\eal}{\end{align}}

\newcommand{\mA}{{\mathsf A}}

\newcommand{\om}{\omega}
\newcommand{\Om}{\Omega}
\newcommand{\tom}{\tilde{\omega}}
\newcommand{\ttom}{\tilde{\tom}}
\newcommand{\tk}{\tilde{k}}

\newcommand{\hu}{\hat u}
\newcommand{\hv}{\hat v}

\newcommand{\Ol}{\overline}

\allowdisplaybreaks

\begin{document}

%\begin{frontmatter}

\title{Second and third harmonic generation of acoustic waves {in a nonlinear elastic solid} in one space dimension}

%% Group authors per affiliation:
\author{Fernando Lund}
\affiliation{Departamento de F\'\i sica and CIMAT, Facultad de Ciencias F\'\i sicas y Matem\'aticas, Universidad de Chile \\ {Av. Blanco Encalada 2008, Santiago, Chile}}
%\fntext[myfootnote]{flund@dfi.uchile.cl}
\email{flund@dfi.uchile.cl}
%% or include affiliations in footnotes:
%\author[mymainaddress,mysecondaryaddress]{Elsevier Inc}
%\ead[url]{www.elsevier.com}

%\author[mysecondaryaddress]{Global Customer Service\corref{mycorrespondingauthor}}
%\cortext[mycorrespondingauthor]{Corresponding author}
%\ead{support@elsevier.com}
%\ead{flund@dfi.uchile.cl}

%\address[mymainaddress]{1600 John F Kennedy Boulevard, Philadelphia}
%\address[mysecondaryaddress]{360 Park Avenue South, New York}

\begin{abstract}
The generation of second and third harmonics by an acoustic wave propagating along one dimension in a weakly nonlinear elastic medium that is loaded harmonically in time with frequency $\om_0$ at a single point in space, is analyzed by successive approximations starting with the linear case. It is noted that nonlinear waves have a speed of propagation that depends on their amplitude, a reflection of the fact that nonlinear oscillators have an amplitude-dependent period. It is also noted that both a free medium as well as a loaded medium generate higher harmonics, but that although the second harmonic of the free medium scales like the square of the linear wave, this is no longer the case when the medium is externally loaded. The shift in speed of propagation due to the nonlinearities is determined imposing that there be no resonant (``secular'') terms in a successive approximations solution scheme to the homogeneous (i.e., ``free'') problem. The result is then used to solve the inhomogeneous (i.e., ``loaded'') case also by successive approximations, up to the third order. At second order, the result is a second harmonic wave whose amplitude is modulated by a long wave, whose wavelength is inversely proportional to the shift in the speed of propagation of the linear wave due to nonlinearities. The amplitude of the long modulating wave scales like the amplitude of the linear wave to the four thirds. It depends both on the third- and fourth-order elastic constants, as well as on the frequency and amplitude of the loading.  At short distances from the source, in a distance scale that depends both on nonlinearities as well as on the loading frequency and on the amplitude of the linear wave, a scaling proportional to the amplitude of the linear wave squared is recovered, as is a second harmonic amplitude that grows linearly with distance from the source, and depends on the third-order elastic constant only. The third order solution is the sum of four amplitude-modulated waves, two of them oscillate with frequency $\om_0$ and the other two, third harmonics, with $3\om_0$. {In each pair, one term scales like the amplitude of the linear wave to the  five-thirds, and the other to the seven-thirds.}
\end{abstract}

%\begin{keyword}
%Nonlinear elastic waves; second harmonic generation.
%\texttt{elsarticle.cls}\sep \LaTeX\sep Elsevier \sep template
%\MSC[2010] 00-01\sep  99-00
%\end{keyword}

%\end{frontmatter}
\maketitle
%\linenumbers
\newpage

\section{Introduction}
Second, and third, harmonic generation by an acoustic wave traveling in a weakly nonlinear elastic solid has become an increasingly useful tool in a variety of solid mechanics contexts. The state of the art up to 2014 was reviewed by Matlack et al. \cite{Matlack2014}. More recently, Achenbach and Wang \cite{Achenbach2018} have studied the generation of higher harmonic surface waves on a half-space of incompressible material. Sahu et al. \cite{Sahu2017} have correlated the generation of second harmonics with the microstructre of P92 steel, and Fuchs et al. \cite{Fuchs2021} have used analytical modeling to relate the formation of precipitates in stainless steel to the behavior of a nonlinearity parameter.  Belotti et al. \cite{Belotti2021} have measured the sensitivity of nonlinear parameters to the presence of dislocations in additively manufactured materials.

Nonlinear parameters appear to be substantially more sensitive to microstructural changes than linear ones, and this has motivated considerable research into the use of nonlinear effects as probes of the microstructure of solid materials. For example, Espinoza et al. \cite{Espinoza2018} noted that second harmonic generation (SHG) is twenty times more sensitive to changes in dislocation density in aluminum than linear Resonant Ultrasound Spectroscopy (RUS), with the added bonus that it is simpler to implement experimentally; and Sosa et al. \cite{Sosa2024} showed, using measuremnts with plastically deformed brass, that measuring the variation in the generation of a second harmonic can potentially discriminate betwee slip and twinning as microscopic mechanisms during plastic deformation. Chakrapani et al. \cite{Chakrapani2017} have related the nonlinearities associated to Nonlinear Resonant Ultrasound Spectroscopy (NRUS) with third order elastic constants. Sahu et al. \cite{Sahu2017} have used nonlinear Rayleigh waves to test the effect of precipitation hardening during tempering of P92 steel. NRUS has been used by Meo et al. \cite{Meo2008} to test for damage in composites of interest to the aerospace industry, and in bone by Muller et al. \cite{Muller2005} and by Haupert et al. \cite{Haupert2015}. Payan et al. \cite{Payan2007,Payan2014} have used nonlinear acoustic techniques to asses damage un concrete. The measurement of second, and higher, harmonics as a way to test for the presence of cracks in glass has been advocated by Persson et al. \cite{Persson2020}. The interest of studying third harmonic generation for non-destructive testing has been highlighted by Kube and Arguelles \cite{Kube2017}, and Lissenden \cite{Lissenden2021} has provided a tutorial reviewing the basic principles underpinning the behavior of nonlinear ultrasonic guided waves from a nondestructive evaluation perspective. The actual measurement of the nonlinear parameters characterizing a solid involve a number of subtleties, a topic that has been reviewed by Jhang et al. \cite{Jhang2020} and Park et al. \cite{Park2021} . Within a geophysical context, the importance of nonlinear elastic behavior has been highlighted by Ostrovsky and Johnson \cite{Ostrovsky2001}. McCall \cite{McCall1994} developed a systematic study of nonlinear elastic wave behavior in an attenuating medium, and quite recently Martin et al. \cite{Martin2019} and O'Brien \cite{OBrien2020} have numerically modeled wave propagation in  nonlinear viscoelastic media, a topic of interest both in seismology and exploration geophysics. De Lima and Hamilton have studied nonlinear waves in plates \cite{deLima2003}. 

A common feature of the analysis used in several of the works mentioned above is to consider that the amplitude $\mA_2$ of the second harmonic and the amplitude  $\mA_1$ of the linear wave measured by a receiver at a distance $x$ from an emitter at angular frequency $\om_0$ in a medium of speed of sound $c_0$ and nonlinear parameter $\beta$ (defined precisely in the next section) are related by  $ \mA_{2} \,    =  \beta \, x \,  \om_0^2 \,  \mA_1^2 /4c_0^2$. It has also been recognized that this solution must be approximate since it increases without bounds away from the emitter, so $x$ must be small. But: small with respect to what? Not with respect to wavelength, since radiation itself is ill-defined at distances from the emitter that are smaller than wavelength. One possibility is to think of attenuation as providing a length scale where $x$ must be small \cite{McCall1994}. However, it still is difficult to imagine how, in an elastic medium without losses, injecting a finite amount of energy per unit time could lead to a diverging energy flux away from the emitter.

A well-known feature of nonlinear oscillators is that their frequency of oscillation depends on amplitude \cite{Nayfeh2008}. One possible way to compute this effect is through a successive approximation scheme starting from the linear case, for small deviations from said linear behavior, imposing that, in the successive approximation scheme resonant, also called secular, terms do not appear. This approach has been used to computed the amplitude-dependent changes in the normal mode frequencies of linear chains of masses joined by nonlinear springs \cite{Swinteck2013,Manktelow2011}. The fact that a continuous elastic medium ought to be the limit of a discrete chain of oscillators in the limit of very small interparticle distance will be used in the results to be obtained below.

In this paper we provide a solution for the one-dimensional propagation of acoustic waves in a non-viscous medium with second and third order nonlinearities in the stress-strain relation that is valid, and bounded, for an arbitrary distance $x$ between emitter and receiver. A length scale is provided by {a combination of} the amplitude of the linear wave, its frequency, the speed of sound and the nonlinear parameters. When $x$ is small with respect to this parameter (Eq. (\ref{eq:condsmallx}) below), the relation mentioned above is recovered. 

\section{The problem}
We wish to solve the equation
\beq
%\boxed{ 
\frac{1}{c_0^2}  \frac{\partial^2 u}{\partial t^2}  - \frac{\partial^2 u}{\partial x^2} = S_0 \delta(x) \sin \om_0t + \beta \frac{\partial}{\partial x}\left( \frac{\partial u}{\partial x} \right)^2  + \gamma \frac{\partial}{\partial x}\left( \frac{\partial u}{\partial x} \right)^3  
%}
\label{eq:fund}
\eeq
for a displacement $u(x,t)$ as a function of position $x$ and time $t$ on the whole real line, $-\infty <x<\infty$,
proceeding by successive approximations, taking the linear equation as the leading order. This assumes that the nonlinear terms introduce small corrections to the linear behavior. Notice that the frequency $\om_0$ introduces a time scale, and together with the speed of sound $c_0$ there is a length scale ($\om_0/c_0$). There is also a dimensionless (source) amplitude that is expected to be small: $S_0 \ll 1$.The dimensionless parameters $\beta$ and $\gamma$, however, are not expected to be small, since they, as well as $c_0$, are material constants, {related to the second, third, and fourth order elastic constants of the nonlinear solid (see for example \cite{McCall1994})} so the success of the approximation scheme rests on gradients of $u$ being small (i.e., small strains): In other words, high wavenumbers will be forbidden. As discussed in the review of Zarembo and Krasil'nikov \cite{Zarembo1971}, it is not possible within this framework{, for example,} to discuss the steepening of wavefronts such as would be associated with shock waves.

\subsection{The continuum problem---linear approximation}
If we ignore the nonlinear terms in (\ref{eq:fund}) we have
\beq
\label{eq:orderone1}
\frac{1}{c_0^2} \frac{\partial^2 u^{(1)}}{\partial t^2} -  \frac{\partial^2 u^{(1)} }{\partial x^2} = S_0 \delta (x) \sin \om_0 t  \, .
\eeq
 This is a linear equation whose solution, for $x \ne 0$, is
\beq
u^{(1)}(x,t) = f(x-c_0t) + g(x+c_0t) \, .
\eeq
We wish to have waves propagating to the right for $x>0$ and to the left for $x<0$. Additionaly, $u(x,t)$ must be continuous at the origin and its first derivative must have a discontinuity at the origin given by
\beq
\left. -\frac{\partial u}{\partial x} \right|_{0^+} + \left. \frac{\partial u}{\partial x} \right|_{0^-} = S_0 \sin \om_0 t
\label{discori}
\eeq
The solution to this problem is
\beq
\label{soleasyone}
u^{(1)}(x,t) = -\frac{c_0 S_0}{2\om_0}\cos \om_0(t- |x|/c_0)   \, .  
\eeq

\subsection{The continuum problem---Fourier transform in space}
 {To continue,} a possible approach (see for example \cite{McCall1994}) is to take Fourier transform of (\ref{eq:fund}) in time and to work in the frequency domain, taking advantage of the fact that the localized-in-$x$ source yields the equation for a well-known Green's function. We will proceed differently, taking Fourier transforms (FT)  in space:
\beq
\label{eq:FTx}
f(x) \equiv \int_{-\infty}^{\infty} \frac{dk}{2\pi} \hat f (k) e^{ikx} \, , \hspace{1em} \hat f(k) \equiv \int_{-\infty}^{\infty} dx f (x) e^{-ikx} \, , \hspace{1em}  \delta (x-x') = \int_{-\infty}^{\infty} \frac{dk}{2\pi} e^{ik(x-x')}
\eeq
and we shall use
\beq
\widehat{\frac{df}{dx}} = ik \hat f(k)
\eeq
as well as
\beq
\widehat{fg} = \hat f * \hat g \, {\equiv \frac{1}{2\pi}} \int_{-\infty}^{\infty} dk' \hat f (k') \hat g(k-k') { = \frac{1}{2\pi} \int_{-\infty}^{\infty} dk' \hat f (k-k') \hat g(k') }
\eeq

So, taking the FT of (\ref{eq:fund}) we have
\bea
\label{basiceq:fkS}
{\frac{1}{c_0^2}}\frac{d^2 \hat u}{d t^2} + k^2  \hat u &=& {S_0 \sin \om_0t  -{\frac{i\beta k}{2\pi} } \int_{-\infty}^{\infty} dk' k' (k-k') \hat u(k') \hat u(k-k') }         \\
&& \hspace{1em} +\frac{\gamma k}{{(2\pi)^2}} \int_{-\infty}^{\infty} dk'  (k-k')  \hat u(k-k')  \int_{-\infty}^{\infty} dk''  (k'-k'') k''  \hat u(k'-k'') \hat u (k'') \, . \nonumber
\eea
Note that the FT of the displacement, $\hat u(k)$, has dimensions of (length)$^2$. Also, the independent variables will be omitted from the equations when the meaning is clear. Eq. \eqref{basiceq:fkS} can be interpreted as a set of nonlinearly coupled harmonic oscillators, labeled by the real number $k$, with linear eigenfrequency $c_0k$, all of them loaded by the same external, {periodic}, force with frequency {$\om_0$} and  {dimensionless amplitude $S_0$}.  

As already mentioned, we shall solve this system by iteration:
\beq
\label{eq:ukap}
\hu(k,t) = \hu^{(1)}(k,t) + \hu^{(2)}(k,t) + \hu^{(3)}(k,t) + \dots 
\eeq In this development we expect to encounter the fact that, for nonlinear waves, the speed of propagation is modified by a term that depends on the amplitude of  the wave. 
More precisely, within this FT context, we will allow the oscillators to respond periodically to the periodic forcing, but with a period that differs from said forcing period. This is a reflection of the fact that the frequency of oscillation of each ($k$-th) oscillator differs from $c_0k$ by an amount that depends on the amplitude of the oscillation. Substituting (\ref{eq:ukap}) into (\ref{basiceq:fkS}) we have
\begin{small}
\bea
&&\hspace{-4em} \frac{1}{c_0^2} \frac{d^2}{dt^2} \left[ \hu^{(1)}(k) + \hu^{(2)}(k) + \hu^{(3)}(k) + \dots \right]   \nonumber \\
&&+ k^2 \left[  \hu^{(1)}(k) + \hu^{(2)}(k) + \hu^{(3)}(k) + \dots \right] 
=  S_0 \sin \om_0t   \label{eq:wnspace}\\
&&{ \hspace{-4em} -\frac{i\beta k}{2\pi}  \int_{-\infty}^{\infty} dk' k' (k-k')  \left[ \hu^{(1)}(k') + \hu^{(2)}(k')  + \dots \right]   \left[ \hu^{(1)}(k-k') + \hu^{(2)}(k-k')  + \dots \right] \nonumber }  \\
&&  \hspace{-3em}  +\frac{\gamma k}{{(2\pi)^2}} \int_{-\infty}^{\infty} dk'  (k-k') [ \hat u^{(1)}(k-k')+\dots ]  \int_{-\infty}^{\infty} dk''  (k'-k'') k''  [\hat u^{(1)}(k'-k'')+\dots ][ \hat u^{(1)} (k'') +\dots ]  \nonumber
\eea
\end{small}

%\vspace{1em}
\subsection{Leading order}
We now look at the leading order problem in wave-number space, Eq. (\ref{eq:wnspace}), with a rewritten right hand side:
\beq
\frac{d^2 \hat u^{(1)}(k)}{c_0^2\, d t^2} + k^2 \hat u^{(1)}(k) =  \frac{S_0}{2i} \left( e^{i\om_0 t +\epsilon t} -e^{-i\om_0 t +\epsilon t} \right)  
\label{eq:loS0}
\eeq
where we take the source to vanish in the remote past ($t\to -\infty$) {by adding a small negative imaginary part to the frequency: $\om_0 \to \om_0 -i\epsilon$}. In going back to position ($x$) space, we will have to take contours in the complex $k$-plane such that $e^{ikx}$ vanishes for large $x$. That is, upper-half-plane for $x>0$, and lower-half-plane for $x<0$, and the position of the poles will be important.

This is a linear problem so we shall have 
\beq
 \hat u^{(1)}(k)  = \hat u^{(1)}_+(k) + \hat u^{(1)}_-(k)
 \label{eq:uonek}
\eeq
with {$\hat u^{(1)}_{\pm}(k)$ the solution of}
\begin{align}
\frac{d^2 \hat u^{(1)}_\pm(k)}{c_0^2\, d t^2} +  k^2 \hat u^{(1)}_\pm(k) =& \pm  \frac{S_0}{2i}  e^{\pm i\om_0 t +\epsilon t}  %\\
\end{align}
 The (steady) solution is (with $\tom_0 \equiv \om_0/c_0$ and we do not distinguish between $\epsilon$ and $\epsilon/c_0$):
\begin{align}
\label{sol:orderonek}
\hat u^{(1)}_\pm(k,t)=& \pm  \frac{ (S_0/2i) e^{\pm i\om_0 t +\epsilon t}}{ k^2 -(\tom_0\mp i\epsilon)^2}  %\\
\end{align}
The general solution would have an additional term, a solution to the homogeneous portion of (\ref{eq:loS0}). We shall assume that this portion has died away, because of viscosity for example.

Note that, according to (\ref{eq:uonek}), we have $\Ol{\hat u^{(1)}(k)} = \hat u^{(1)}(-k)$ so it is the FT of a real function. Note also that expression (\ref{sol:orderonek}) has a singularity at $k= \pm \om_0/c_0$. This means that the equation picks a precise wave vector for a given loading frequency: of all the available ``$k$'' oscillators, only one responds significantly. This relationship between wave vector and loading frequency determines the speed of propagation of the wave when going back to position space.

We now use Cauchy's theorem:
\beq
\oint \frac{ f(k)dk}{k-k_0} = 2\pi i f(k_0)
\eeq
with the contour integral going counterclockwise.
For $x>0$,
\bea
u^{(1)}_+(x,t) & = &-  \frac{c_0S_0}{4\om_0} e^{i\om_0 (t-x/c_0)}  %\\
\eea
so we recover (\ref{soleasyone}) for $x>0$. And, for arbitrary $x$,
\bea
u^{(1)}_+(x,t) & = &-  \frac{c_0S_0}{4\om_0} e^{i\om_0 (t-|x|/c_0)}  %\\
\eea
so that
\beq
u^{(1)}(x,t) = -  \frac{c_0S_0}{2\om_0} \cos \om_0 (t-|x|/c_0) 
\label{eq:sol_o1}
\eeq
as it should.

In order to continue to higher orders that take into account the nonlinear coupling among oscillators we remember that nonlinear oscillators  have a frequency that differs from the linear frequency, by an amount that depends on the nonlinear parameters. We shall now address this problem. First the homogeneous problem will be solved in order to determine the shift in frequency due to the nonlinear behavior---through the imposition that resonances be absent---of the oscillator. Ref. \cite{Khoo1976} can be used  as a guide. Then the so determined shift {will be} plugged into the response to an external loading. 

Notice that if $u(x,t)$ is a solution of (\ref{eq:fund}) for $x>0$, then the solution for $x<0$ is obtained from the fact that $u(-x,t)$ is a solution of the same equation with $\beta$ changed to $-\beta$.

\section{The homogeneous problem}
We consider then the homogeneous portion of Eqn. (\ref{eq:fund})
\beq
\frac{1}{c_0^2}  \frac{\partial^2 v}{\partial t^2}  - \frac{\partial^2 v}{\partial x^2} = \beta  \frac{\partial}{\partial x} \left( \frac{\partial v}{\partial x} \right)^2 + \gamma  \frac{\partial}{\partial x} \left( \frac{\partial v}{\partial x} \right)^3
\eeq
whose FT is
\bea
\label{eq:homknodisp}
{\frac{1}{c_0^2}}\frac{d^2 \hv}{d t^2} + k^2  \hv &=&  -{\frac{i\beta k}{2\pi} } \int_{-\infty}^{\infty} dk' k' (k-k') \hv(k') \hv(k-k')    \\
&& \hspace{2em} +\frac{\gamma k}{{(2\pi)^2}} \int_{-\infty}^{\infty} dk'  (k-k')  \hv(k-k')  \int_{-\infty}^{\infty} dk''  (k'-k'') k''  \hv(k'-k'') \hv (k'')  \nonumber   
\eea

Introducing (notice the slight change in notation {compared to (\ref{eq:ukap})}) 
\beq
\hv (k ) =  \hv^{(0)}(k) + \hv^{(1)}(k) + \hv^{(2)}(k) + \dots
\label{successiveapproxhom}
\eeq
we have
\begin{small}
\bea
&&\hspace{-4em} \frac{1}{c_0^2} \frac{d^2}{dt^2} \left[ \hv^{(0)}(k) + \hu^{(1)}(k) + \hv^{(2)}(k) + \dots \right]   \nonumber \\
&&+ k^2 \left[  \hv^{(0)}(k) + \hv^{(1)}(k) + \hv^{(2)}(k) + \dots \right] 
=    \label{eq:wnspacehom}\\
&&{ \hspace{-2em} -\frac{i\beta k}{2\pi}  \int_{-\infty}^{\infty} dk' k' (k-k')  \left[ \hv^{(0)}(k') + \hv^{(1)}(k')  + \dots \right]   \left[ \hv^{(0)}(k-k') + \hv^{(1)}(k-k')  + \dots \right] \nonumber }   \\
&&  +\frac{\gamma k}{2\pi} \int_{-\infty}^{\infty} dk'  (k-k') [ \hat v^{(0)}(k-k') + \dots ] \int_{-\infty}^{\infty} dk''  (k'-k'') k''  [\hat v^{(0)}(k'-k'') + \dots ] [ \hat v^{(0)} (k'') + \dots ] \nonumber \\
\eea
\end{small}
whose leading order 
\beq
\frac{1}{c_0^2} \frac{d^2 \hv^{(0)}(k)}{dt^2} + k^2  \hv^{(0)}(k) = 0
\eeq%Ff
has the solution 
\begin{align}
 \hv^{(0)}(k) = & \, A_k^{(0)} e^{i(\om^{(0)}_k + \om^{(1)}_k + \om^{(2)}_k + \dots ) t } + {B_k^{(0)} e^{-i(\om^{(0)}_k + \om^{(1)}_k + \om^{(2)}_k + \dots ) t }} \\
  \equiv & \, \hv^{(0)}_+(k) + {\hv^{(0)}_{-}(k)}
 \label{u+okhom}
 \end{align}
 with $\om^{(0)}_k \equiv c_0k$, {and we shall let $\om^{(i)}_k = c_i k; i=1,2 \dots$}. {Also $\Ol{A_k^{(0)}} = A_{-k}^{(0)}$ and $\Ol{B_k^{(0)}} = B_{-k}^{(0)}$. }We have allowed for a frequency of oscillation that deviates form the linear case, with corrections that may be $k$ dependent, and {that} will depend on the amplitudes $A_k^{(0)}$ and $B_k^{(0)}$. These amplitudes also can depend on $k$, and are determined by initial conditions.
 
The reasoning is that the right-hand-side of (\ref{eq:wnspacehom}) should not give rise to terms that have a time dependence {proportional to} $e^{\pm ic_0kt}$ because they would induce resonances whose amplitude increase without bound. The hope is that imposing this condition on the iteration will yield values for the frequency corrections $\om^{(1)}_k, \om^{(2)}_k.  \dots$. 

 \subsection{Order one}
This is
\begin{small}
\bea
\hspace{-2em} \frac{1}{c_0^2} \frac{d^2 \hv^{(1)}(k)}{dt^2} + k^2  \hv^{(1)}(k) &=& -\frac{i\beta k}{2\pi}  \int_{-\infty}^{\infty} dk' k' (k-k') \hv^{(0)}(k') \hv^{(0)}(k-k') + \frac{2\om^{(0)}_k \om^{(1)}_k}{c_0^2} \hv^{(0)}(k)    \nonumber \\
& = &  -\frac{i\beta k}{2\pi}  \int_{-\infty}^{\infty} dk' k' (k-k') \left(   \hv^{(0)}_+(k') + {\hv^{(0)}_-(k')} \right) \left(  \hv^{(0)}_+(k-k') + {\hv^{(0)}_-(k-k')} \right)    \nonumber \\
   && \hspace{4em}  + \frac{2\om^{(0)}_k \om^{(1)}_k}{c_0^2} \hv^{(0)}(k) 
\label{orderoneinhom}
\eea
\end{small}
whose right-hand-side (RHS) has terms that, in an  obvious notation, we shall call ``$++$'', ``$+-$'', ``$-+$'' and ``$--$''. We take ``$++$'' first:
\bea
RHS_{++} &\equiv& -\frac{i\beta k}{2\pi}  \int_{-\infty}^{\infty} dk' k' (k-k') {\hv^{(0)}_-(k')} \, {\hv^{(0)}_-(k-k')}   \\
&=&  -\frac{i\beta k}{2\pi}  \int_{-\infty}^{\infty} dk' k' (k-k') A_{k^{'}}^{(0)} A_{k-k^{'}}^{(0)}  \\
&& \hspace{1em} \times \, e^{i(\om^{(0)}_{k^{'}} + \om^{(1)}_{k^{'}} + \om^{(2)}_{k^{'}} + \dots ) t} e^{i(\om^{(0)}_{k-k^{'}} + \om^{(1)}_{k-k^{'}} + \om^{(2)}_{k-k^{'}} + \dots ) t }  \nonumber
\eea
Similarly, 
\bea
RHS_{--} &\equiv& -\frac{i\beta k}{2\pi}  \int_{-\infty}^{\infty} dk' k' (k-k') \hv^{(0)}_+(k') \hv^{(0)}_+(k-k')   \\
&=& -\frac{i\beta k}{2\pi}  \int_{-\infty}^{\infty} dk' k' (k-k') B_{k^{'}}^{(0)} B_{k-k^{'}}^{(0)}  \\
&& \hspace{1em} \times \, e^{-i(\om^{(0)}_{k^{'}} + \om^{(1)}_{k^{'}} + \om^{(2)}_{k^{'}} + \dots ) t} e^{-i(\om^{(0)}_{k-k^{'}} + \om^{(1)}_{k-k^{'}} + \om^{(2)}_{k-k^{'}} + \dots ) t } \, . \nonumber  
\label{eq:homRHS++}
\eea
If we were to allow, as would seem natural,
\beq
 \om^{(0)}_k  = c_0k
 \label{w=c0k}
\eeq
we would have  $\om^{(0)}_{k^{'}} + \om^{(0)}_{k-k^{'}} = \om^{(0)}_{k}$ independently of $k'$ and would conclude that $RHS_{++}$ and $RHS_{--}$ give rise to ``resonant'' terms at leading order. In order to avoid this contradictory behavior we remember that the problem at hand can be considered as the limit of a chain of discrete nonlinear oscillators, for very small inter-oscillator distance compared to wavelength. This discrete problem has been studied by Swinteck et al. \cite{Swinteck2013}, and we shall lean on it to develop a consistent approach to the continuum problem. For current purposes, the main feature of the discrete problem is the existence of a length scale ``$a$'', the separation between masses in equilibrium. Its existence introduces dispersion effects, that is, the frequency of oscillation is no longer proportional to wave vector as in (\ref{w=c0k}) but is replaced by $ \om^{(0)}_k  \to (2c_0/a) \sin (ka/2)$. In order to implement these facts we replace (\ref{eq:homknodisp}) by
\bea
\label{homprob6latt}
{\frac{a^2}{{c_0^2}}}\frac{d^2 \hat v}{d t^2} + K_k^2  \hat v &=&  -{\frac{i\beta ka^2}{2\pi} } \int_{-\infty}^{\infty} dk' k' (k-k') \hat v(k') \hat v(k-k')    \\
&& \hspace{2em} +\frac{\gamma ka^2}{{(2\pi)^2}} \int_{-\infty}^{\infty} dk'  (k-k')  \hat v(k-k')  \int_{-\infty}^{\infty} dk''  (k'-k'') k''  \hat v(k'-k'') \hat v (k'')  \nonumber   
\eea
where we have introduced the dimensionless wavenumber
\beq
K_k \equiv {2\sin\left(\frac{ka}{2}\right)}    = ak - \frac 1{24} (ak)^3 + \cdots
\eeq
which would be the case if the propagation was on a lattice of spacing ``$a$''  rather than on a continuum {(see \cite{Swinteck2013} for the details of propagtion on a lattice)}. 
Equation (\ref{homprob6latt}) reduces to (\ref{eq:homknodisp}) in the limit $ka \to 0.$

Introducing the successive approximations (\ref{successiveapproxhom}) into this expressions we get 
\begin{small}
\bea
&&\hspace{-4em} \frac{a^2}{{c_0^2}} \frac{d^2}{dt^2} \left[ \hv^{(0)}(k) + \hv^{(1)}(k) + \hv^{(2)}(k) + \dots \right]   \nonumber \\
&&+ K_k^2 \left[  \hv^{(0)}(k) + \hv^{(1)}(k) + \hv^{(2)}(k) + \dots \right] 
=    \label{eq:wnspacehom}\\
&&{ \hspace{-2em} -\frac{i\beta ka^2}{2\pi}  \int_{-\infty}^{\infty} dk' k' (k-k')  \left[ \hv^{(0)}(k') + \hv^{(1)}(k')  + \dots \right]   \left[ \hv^{(0)}(k-k') + \hv^{(1)}(k-k')  + \dots \right] \nonumber }   \\
&&  +\frac{\gamma ka^2}{{(2\pi)^2}} \int_{-\infty}^{\infty} dk'  (k-k') [ \hat v^{(0)}(k-k') + \dots ] \int_{-\infty}^{\infty} dk''  (k'-k'') k''  [\hat v^{(0)}(k'-k'') + \dots ] [ \hat v^{(0)} (k'') + \dots ] \nonumber \\
\eea
\end{small}
whose leading order 
\beq
\frac{a^2}{{c_0^2}} \frac{d^2 \hv^{(0)}(k)}{dt^2} + K_k^2  \hv^{(0)}(k) = 0
\label{hom:lo}
\eeq
has the solution 
\bea
\label{eq:homsol0}
 \hv^{(0)}(k) &=& A_k^{(0)} e^{i(\Om^{(0)}_k + \Om^{(1)}_k + \Om^{(2)}_k + \dots ) t} + {B_k^{(0)} e^{-i(\Om^{(0)}_k + \Om^{(1)}_k + \Om^{(2)}_k + \dots ) t}} \\
 & \equiv & \hv^{(0)}_+(k) + {\hv^{(0)}_{-}(k)}
 \label{u+okhom}
 \eea
 with $\Om^{(0)}_k \equiv c_0K_k/a$ and $\Om^{(i)}_k = c_i K_k/a; i=1,2 \dots$. We have allowed for a frequency of oscillation that deviates from the linear case, with corrections that may be $k$ dependent, and will depend on the amplitudes $A_k^{(0)}$ and $B_k^{(0)}$.

\newpage
The order one equation is
\begin{small}
\bea
\label{o1homsimple}
\hspace{-2em} \frac{a^2}{{c_0^2}} \frac{d^2 \hv^{(1)}(k)}{dt^2} + K_k^2  \hv^{(1)}(k) &=& -\frac{i\beta ka^2}{2\pi}  \int_{-\infty}^{\infty} dk' k' (k-k') \hv^{(0)}(k') \hv^{(0)}(k-k') + \frac{2\Om^{(0)}_k \Om^{(1)}_k}{c_0^2/a^2} \hv^{(0)}(k)     \\
& = &  -\frac{i\beta ka^2}{2\pi}  \int_{-\infty}^{\infty} dk' k' (k-k') \left(   \hv^{(0)}_+(k') + {\hv^{(0)}_-(k')} \right) \left(  \hv^{(0)}_+(k-k') + {\hv^{(0)}_-(k-k')} \right)    \nonumber \\
   && \hspace{4em}  + \frac{2\Om^{(0)}_k \Om^{(1)}_k}{c_0^2/a^2} u^{(0)}(k)
\label{orderoneinhOm}
\eea
\end{small}
which has the same structure as (\ref{orderoneinhom}).
%\subsubsection{A simpler reasoning}
Take a closer look at (\ref{o1homsimple}): it is a harmonic oscillator driven by the terms on the right-hand-side. The first term on the right is a bilinear combination of the order zero oscillator: Infinitely many of them combine to drive the $k-th$ order one oscillator. But not all of them will contribute to a resonant driving. Indeed, since $u^{(0)}(k)$ is a linear combination of oscillators with frequencies $\pm\Om_k$, the only way for the bilinear combination to give rise to a resonance is to have either $k'=k$, or $k'=0$. But in both cases the kernel of the integrand vanishes. The conclusion is that this bilinear combination does not lead to secular terms and, consequently, 
\beq
\Om^{(1)}_k = 0 
\eeq

 In order go to the next order of approximation we need the solution to 
 \beq
\frac{a^2}{{c_0^2}} \frac{d^2 \hv_{++}^{(1)}(k)}{dt^2} + K_k^2  \hv_{++}^{(1)}(k) = RHS_{++}
\label{oone++}
\eeq
which is 
\beq
\hv_{++}^{(1)}(k) = \frac{i\beta ka^2}{2\pi}  \int_{-\infty}^{\infty} dk'  k' (k-k') A_{k^{'}}^{(0)} A_{k-k^{'}}^{(0)}  \frac{e^{i(\Om_{k-k'} +\Om_{k'}) t} }{\left( K_{k-k'} +K_{k'}\right)^2 -K_k^2} \, .
\eeq
A similar reasoning leads to 
\beq
\hv_{--}^{(1)}(k) = \frac{i\beta ka^2}{2\pi}  \int_{-\infty}^{\infty}dk' k' (k-k') B_{k^{'}}^{(0)} B_{k-k^{'}}^{(0)}  \frac{e^{-i(\Om_{k-k'} +\Om_{k'}) t} }{\left( K_{k-k'} +K_{k'}\right)^2 -K_k^2}
\eeq
\beq
\hv_{+-}^{(1)}(k) =   \frac{i\beta ka^2}{2\pi}  \int_{-\infty}^{\infty} dk'  k' (k-k') A_{k^{'}}^{(0)} B_{k-k^{'}}^{(0)} \frac{e^{i (\Om_{k'} -\Om_{k-k^{'}})t} }{\left( K_{k^{'}} -K_{k -k^{'}} \right)^2 -K_k^2} 
\label{eq92}
\eeq
and the order-one solution to the homogeneous problem is (ignoring the possibility of solutions to the homogeneous equation)
\beq
\label{eq:uone}
\hv^{(1)}(k) = \, \hv^{(1)}_{++}(k) + \hv^{(1)}_{--}(k) + 2\hv^{(1)}_{+-}(k) 
\eeq

\subsection{Order two}
This is 
\begin{small}
\bea
\label{otwohomone}
\frac{a^2}{c_0^2} \frac{d^2 \hv^{(2)}(k)}{dt^2} + K_k^2  \hv^{(2)}(k) &=& -\frac{i\beta ka^2}{2\pi}  \int_{-\infty}^{\infty} dk' k' (k-k')[ \hv^{(1)}(k') \hv^{(0)}(k-k') + \hv^{(0)}(k') \hv^{(1)}(k-k') ]  \nonumber \\
&& \hspace{-2em} +\frac{\gamma k a^2}{{(2\pi)^2}} \int_{-\infty}^{\infty} dk'  (k-k')  \hat v^{(0)}(k-k')  \int_{-\infty}^{\infty} dk''  (k'-k'') k''  \hat v^{(0)}(k'-k'') \hat v^{(0)} (k'')  \nonumber \\
&& \hspace{2em} + \frac{2\Om^{(0)}_k \Om^{(2)}_k}{c_0^2/a^2}  \hv^{(0)}(k) \nonumber \\[0.5 em]
&\equiv & RHS^{\beta}  + RHS^{\gamma} +  \frac{2\Om^{(0)}_k \Om^{(2)}_k}{c_0^2/a^2} \hv^{(0)}(k)
\label{otwohom}
\eea
\end{small}
and we must identify the possible resonant terms in the RHS. 
 
 \subsubsection{Beta}
 \label{subsubsecbeta}
We shall now take the first order solution that was obtained as a ``limit'' of a discrete chain. Consider
\bea
\label{rhsbetauf}
RHS^{\beta}  &\equiv& -\frac{i\beta ka^2}{\pi}  \int_{-\infty}^{\infty} dk' k' (k-k') \hv^{(1)}(k') \hv^{(0)}(k-k')  \\
&=&  \frac{\beta^2 ka^4}{2\pi^2}  \int_{-\infty}^{\infty} dk' {k'}^2 (k-k') \left[ A_{k-k'}^{(0)} e^{i\Om_{k-k'}t} + {B_{k-k'}^{(0)}} e^{-i\Om_{k-k'}t} \right]   \nonumber \\
&& \times \left[   \int_{-\infty}^{\infty} dk''  k'' (k'-k'') A_{k^{''}}^{(0)} A_{k'-k^{''}}^{(0)}  \frac{e^{i(\Om_{k'-k''} +\Om_{k''}) t} }{\left( K_{k'-k''} +K_{k''}\right)^2 -K_{k'}^2}   \right. \nonumber \\
&& \hspace{1em} +    \int_{-\infty}^{\infty}dk'' k'' (k'-k'') B_{k^{''}}^{(0)} B_{k'-k^{''}}^{(0)}  \frac{e^{-i(\Om_{k'-k''} +\Om_{k''}) t} }{\left( K_{k'-k''} +K_{k''}\right)^2 -K_{k'}^2}             \nonumber \\ 
&& \hspace{1em} +2 \left.    \int_{-\infty}^{\infty} dk''  k'' (k'-k'') A_{k^{''}}^{(0)} B_{k'-k^{''}}^{(0)} \frac{e^{i (\Om_{k''} -\Om_{k'-k^{''}})t} }{\left( K_{k^{''}} -K_{k' -k^{''}} \right)^2 -K_{k'}^2}   \right] \\
& \equiv & RHS^{\beta}_{+++} + RHS^{\beta}_{+--} + RHS^{\beta}_{-++} + RHS^{\beta}_{---} + RHS^{\beta}_{++-} +  RHS^{\beta}_{-+-}    
\label{betasix}
\eea
As noted, this is the sum of six terms. We are looking for those terms whose time dependence is given by $e^{\pm i\Om_kt}$ only: It is they that lead to resonant behavior.

To expand a little bit: Eqn. (\ref{otwohom}) represents a single harmonic oscillator, $\hv^{(2)}(k)$, labeled by $k$, driven by a bilinear sum of many oscillators, labeled by  $k, k^{'}$ and $k^{''}$. The generic terms in this sum do not lead to resonant behavior of $\hv^{(2)}(k)$, but a few might. And the task at hand is to identify them, remembering that we are interested in the limit $ka \to 0$, where the following behavior will be useful:
\bea
\label{limitone}
\left( K_{k'-k} +K_{k}\right)^2 -K_{k'}^2 & \longrightarrow & \frac 1{8} {k'}^2 k (k'-k) a^4   \\
 \left( K_{k} -K_{k' -k} \right)^2 -K_{k'}^2 & \longrightarrow & 4k(k-k')a^2  \label{limittwo}\\
 & \mbox{\small $ka \to 0$} & \nonumber
\eea

We start by considering the first four terms in the right-hand-side of (\ref{betasix}):
\begin{align}
RHS^{\beta}_{+++} + RHS^{\beta}_{+--} &+ RHS^{\beta}_{-++} + RHS^{\beta}_{---}  \\
 = & \,  \frac{\beta^2 ka^4}{2\pi^2}  \int_{-\infty}^{\infty} dk' {k'}^2 (k-k') \left[ A_{k-k'}^{(0)} e^{i\Om_{k-k'}t} + {B_{k-k'}^{(0)}} e^{-i\Om_{k-k'}t} \right]   \nonumber \\
& \, \left[   \int_{-\infty}^{\infty} dk''  k'' (k'-k'') A_{k^{''}}^{(0)} A_{k'-k^{''}}^{(0)}  \frac{e^{i(\Om_{k'-k''} +\Om_{k''}) t} }{\left( K_{k'-k''} +K_{k''}\right)^2 -K_{k'}^2 }   \right. \nonumber \\
& \hspace{1em} +  \left.  \int_{-\infty}^{\infty}dk'' k'' (k'-k'') B_{k^{''}}^{(0)} B_{k'-k^{''}}^{(0)}  \frac{e^{-i(\Om_{k'-k''} +\Om_{k''}) t} }{\left( K_{k'-k''} +K_{k''}\right)^2 -K_{k'}^2}       \right]   \,    \nonumber
\end{align}
In order for these terms to lead to resonant behavior one of two things must happen: Either
\beq
\Om_{k-k'} - \Om_{k'-k''} -\Om_{k''} =\pm \Om_k
\label{poss1}
\eeq
or
\beq
\Om_{k-k'} + \Om_{k'-k''} +\Om_{k''} =\pm \Om_k
\label{poss2}
\eeq
The first possibility, Eqn. (\ref{poss1}) can only be satisfied when one of $k', k'', (k-k')$ or $(k'-k'')$ vanish. But in this case the kernel of the corresponding integral vanishes as well and these oscillators turn out to be harmless. The second  possibility, Eqn. (\ref{poss2}) has, however, a nontrivial realization, namely $k''=k$. The contribution of this oscillator to the right-hand-side of (\ref{betasix}) is, remembering that $\Om_k$ is an odd function of $k$ and that $A^{(0)}_{-k} =\Ol{A^{(0)}_{k}}$,
\begin{small}
\begin{align}
RHS^{\beta}_{+++} + RHS^{\beta}_{+--} &+ \left. RHS^{\beta}_{-++} + RHS^{\beta}_{---} \right|_{k''=k}  \\
 = &\, -\frac{\beta^2 k^2a^4}{2\pi^2}  \int_{-\infty}^{\infty} dk' \frac{{k'}^2 (k-k')^2}{\left( K_{k'-k} +K_{k}\right)^2 -K_{k'}^2}   \nonumber \\
 & \times \left[ A_{k}^{(0)} \left| A_{k'-k}^{(0)} \right|^2 e^{i\Om_{k} t}   +A_{k-k'}^{(0)} B_{k^{}}^{(0)} B_{k'-k}^{(0)} e^{-i(2\Om_{k'-k} +\Om_{k}) t} \right. \nonumber \\
 &  \qquad + \left. {B_{k-k'}^{(0)}}A_{k}^{(0)} A_{k'-k}^{(0)}e^{i(2\Om_{k'-k} +\Om_{k}) t} + \left| {B_{k-k'}^{(0)}} \right|^2 B_{k^{}}^{(0)} e^{-i\Om_{k} t} \right]  \nonumber \\
 \longrightarrow & \, \frac{{4}\beta^2 k}{\pi^2}  \int_{-\infty}^{\infty} dk' (k-k') \left[ \left| A_{k'-k}^{(0)} \right|^2 A_{k}^{(0)}e^{i\Om_{k} t} + \left| B_{k'-k}^{(0)} \right|^2 B_{k^{}}^{(0)} e^{-i\Om_{k} t} \right] \left[   +   \right] + HT
 \label{RHSo2a}
\end{align}
\end{small}
where, in the last line, ``$HT$'' means ``harmless terms'', that is, terms that do not give rise to resonances, and the limit $ka \to 0$ has been taken, using (\ref{limitone}). And this last integral vanishes because the integrand is an odd function. The ``$ABB$'' and ``$BAA$'' terms are harmless terms ($HT$) because the only way for them to be resonant would be to have $k' =0$ or $k'=k$, and in both cases the integral's kernel vanishes. It is assumed that $A_{k}^{(0)}$ and $B_{k^{}}^{(0)}$ vanish when the condition $ka \ll 1$ is not satisfied.

We are thus left with 
\begin{small}
\bea
RHS^{\beta}  &=& \frac{\beta^2 ka^4}{\pi^2}  \int_{-\infty}^{\infty} dk' {k'}^2 (k-k') \left[ A_{k-k'}^{(0)} e^{i\Om_{k-k'}t} + {B_{k-k'}^{(0)}} e^{-i\Om_{k-k'}t} \right]     \nonumber \\
&& \times  \int_{-\infty}^{\infty} dk''  k'' (k'-k'') A_{k^{''}}^{(0)} B_{k'-k^{''}}^{(0)} \frac{e^{i (\Om_{k''} -\Om_{k'-k^{''}})t} }{\left( K_{k^{''}} -K_{k' -k^{''}} \right)^2 -K_{k'}^2}   + HT \nonumber \\
& \longrightarrow & - \frac{\beta^2 ka^2}{4\pi^2}  \int_{-\infty}^{\infty} dk' {k'}^2 (k-k') \left[ A_{k-k'}^{(0)} e^{i\Om_{k-k'}t} +{B_{k-k'}^{(0)}} e^{-i\Om_{k-k'}t} \right]     \nonumber \\
&& \times \int_{-\infty}^{\infty} dk'' A_{k^{''}}^{(0)} B_{k'-k^{''}}^{(0)} e^{i (\Om_{k''} -\Om_{k'-k^{''}})t} + HT
\eea
\end{small}
where the limit $ka \to 0$ has been taken. The ``$AAB$'' term has a resonance when $k''=k$ and $k'=2k$. Indeed, in that case
\bea
\Om_{k-k'} + \Om_{k''} -\Om_{k'-k^{''}}& \to&  -\Om_k \\
\mbox{\rm and} \qquad A_{k-k'}^{(0)} A_{k^{''}}^{(0)} B_{k'-k^{''}}^{(0)} &\to& \left| A_{k}^{(0)} \right|^2 B_k^{(0)}
\eea
and the ``$BAB$'' term, when $k'-k''=k$ and $k'=2k$ since
\bea
-\Om_{k-k'} + \Om_{k''} -\Om_{k'-k^{''}}& \to&  \Om_k \\
\mbox{\rm and} \qquad B_{k-k'}^{(0)} A_{k^{''}}^{(0)} B_{k'-k^{''}}^{(0)} &\to& \left| B_{k}^{(0)} \right|^2 A_k^{(0)}
\eea

So we finally have
\bea
\left. RHS^{\beta}  \right|_{\rm reso} &=& \frac{\beta^2 k^4a^2}{\pi^2}   \left[   A_{k}^{(0)} e^{i\Om_{k}t} \int_{k' \sim 2k} dk' \int_{k'' \sim k} dk'' {B_{k-k'}^{(0)}} B_{k^{''}}^{(0)} \right. \nonumber \\ 
&& \hspace{4em} \left. + B_{k}^{(0)} e^{-i\Om_{k}t} \int_{k' \sim 2k} dk' \int_{k'' \sim k} dk'' A_{k-k'}^{(0)} A_{k^{''}}^{(0)} \right]   
\label{rhsbetareso} 
\eea
and for the moment we do not need to be precise about what it means to integrate in the neighborhood of a certain value only, as in $k' \sim 2k$. However, as it will become apparent in the next section, due to the inhomogeneous problem we are trying to solve we are particularly interested in 
\beq
\label{only2}
\hat v^{(0)}(k,t)= {\cal A}  \left[ \delta (\tk -1) - \delta ( \tk +1) \right] \left[ e^{i\Om_kt} -  e^{-i\Om_k t} \right] \, ,
\eeq
that is
\beq
A_k^{(0)} =  {\cal A}  \left[ \delta (\tk -1) - \delta ( \tk +1) \right]  = -B_k^{(0)}  \, .
\eeq
with ${\cal A}$ a pure imaginary number and $\tk \equiv k/k_0$ with $k_0$ a fixed real number with dimensions of wave vector. That is, of the infinitely many oscillators available, only two, $k=\pm k_0$, are active at lowest order. In this case \eqref{rhsbetareso} becomes
\beq
\left. RHS^{\beta}  \right|_{\rm reso} = \frac{\beta^2 k^4a^2}{\pi^2}  \hat v^{(0)}(k,t) \, {\cal J} (k)
\eeq
with
\begin{align}
{\cal J} (k) \equiv & \int_{k' \sim 2k} dk' \int_{k'' \sim k} dk'' A_{k-k'}^{(0)} A_{k^{''}}^{(0)} \\
 = &\,   {\cal A}^2 \int_{k' \sim 2k} dk' \int_{k'' \sim k} dk''  \left[ \delta (\tk -\tk' -1) - \delta ( \tk -\tk' +1) \right]  \left[ \delta (\tk'' -1) - \delta ( \tk'' +1) \right] 
\end{align}

We now have to be precise about ``$k' \sim 2k$''. What we really mean is $k'=2k$, and we are interested in
\begin{align}
M(k) \equiv & \, \int_{k'=2k} d\tk' \, \delta(\tk -\tk' -1) = 
\left\{
\begin{array}{ll}
1 & \mbox{if $\tk = -1$} \\
0 & \mbox{if $\tk \ne -1$}
\end{array}               
\right. \\
N(k) \equiv & \, \int_{k'' = k} dk'' \, \delta (\tk'' -1) = 
\left\{
\begin{array}{ll}
1 & \mbox{if $\tk = 1$} \\
0 & \mbox{if $\tk \ne 1$}
\end{array}               
\right.
\end{align}
so that
\beq
{\cal J}(k) = {\cal A}^2 k_0^2 \left[ M(k) - M(-k) \right]\left[ N(k) -N(-k) \right]   
\eeq
and
\beq
\left. RHS^{\beta}  \right|_{\rm reso} = - \frac{2\beta^2 k_0^6a^2{\cal A}^2}{\pi^2}  \hat v^{(0)}(k,t) \, 
\eeq

\subsubsection{Gamma}
Now we take 
%\begin{small}
\beq
RHS^{\gamma} = \frac{\gamma k {a^2}}{{(2\pi)^2}} \int_{-\infty}^{\infty} dk'  (k-k')  \hat v^{(0)}(k-k')  \int_{-\infty}^{\infty} dk''  (k'-k'') k''  \hat v^{(0)}(k'-k'') \hat v^{(0)} (k'')   
\eeq
%\end{small}
and reason as in Section \ref{subsubsecbeta}: the right-hand-side is a trilinear combination of oscillators $\hat u^{(0)}(k)$ oscillating with frequencies $\Om_k$, and  only a subset in the sum will lead to resonant behavior. Indeed, it is necessary that either $k'' =k$,  or $k'=0$ (notice that, apparently additional, case $k'-k'' = k$ is the same as $k''=k$ since the last integral is a convolution). Take $k''=k$ first, and then $k'=0$. This gives
\beq
\label{gammaresdd}
 \left. RHS^{\gamma} \right|_{k''=k} = -\frac{\gamma k^2{a^2}}{{(2\pi)^2}}  {\cal K}(k,t)   \hat v^{(0)} (k) = \left. RHS^{\gamma} \right|_{k'=0} \, .
\eeq
with
\beq
{\cal K}(k,t) \equiv \int_{-\infty}^{\infty} dk'  (k-k')^2  \hat v^{(0)}(k-k')  \int_{k'' \sim k} dk''   \hat v^{(0)}(k'-k'')
\eeq
so that
\beq
\left. RHS^{\gamma} \right|_{\rm reso} = 2 \left. RHS^{\gamma} \right|_{k''=k}
\eeq

In the special case \eqref{only2} we are interested this gives
\begin{align}
 \int_{k'' \sim k} dk''   \hat u^{(0)}(k'-k'') = & \, {\cal A}\, k_0  \int_{k'' = k} d\tk'' \left[ \delta (\tk' - \tk'' -1) - \delta ( \tk'-\tk'' +1) \right] \left[ e^{i\Om_{k'-k''}t} -  e^{-i\Om_{k'-k''} t} \right] \nonumber  \\
 = & \, 2{\cal A}k_0 \left[ e^{i\Om_{k_0}t} -  e^{-i\Om_{k_0} t} \right] 
\end{align}
so that 
\begin{align}
{\cal K}(k,t) = & \, 2{\cal A}^2 k_0^4 \int_{-\infty}^{\infty} d\tk'  (\tk-\tk')^2 \left[ \delta (\tk - \tk' -1) - \delta ( \tk-\tk' +1) \right]  \nonumber \\
& \hspace{4em} \times \left[ e^{i\Om_{k-k'}t} -  e^{-i\Om_{k-k'} t} \right] \left[ e^{i\Om_{k_0}t} -  e^{-i\Om_{k_0} t} \right] \nonumber \\
= & \, 4 {\cal A}^2 k_0^4 \left[ e^{2i\Om_{k_0}t} -2 +  e^{-2i\Om_{k_0} t} \right]
\end{align}
and, from \eqref{gammaresdd},
\beq
\left. RHS^{\gamma} \right|_{\rm reso} = -\frac{2\gamma k_0^6{a^2}{\cal A}^2}{{\pi^2}}  \hat v^{(0)} (k)
\eeq

Putting together tems in (\ref{otwohom}) we have  
\bea
\label{O2hom}
0&=&  \left. RHS^{\beta}\right|_{\rm reso}  + \left. RHS^{\gamma} \right|_{\rm reso}+  \frac{2\Om^{(0)}_k \Om^{(2)}_k}{c_0^2/a^2} \hv^{(0)}(k) \\
&=&  - \frac{2\beta^2 k_0^6a^2{\cal A}^2}{\pi^2}  \hat v^{(0)}(k,t)  -\frac{2\gamma k_0^6{a^2}{\cal A}^2}{{\pi^2}}  \hat v^{(0)} (k)  +\frac{2\Om^{(0)}_k \Om^{(2)}_k}{c_0^2/a^2} \hv^{(0)}(k) 
\eea
which gives
\beq
     \frac{c_2}{c_0} =     \frac{k_0^4 {\cal A}^2}{{\pi^2}} \left[ {}\gamma  + \beta^2  \right]  \, .
\label{finalc2}
\eeq

\section{Back to the inhomogeneous case}
In order to deal with the inhomogeneous case we revisit (\ref{eq:wnspace}), and note that  now not all the oscillators labeled by $k$ are equivalent: two of them, the ones with $k=\pm \tom_0$, are driven at resonance. Hence they will not respond by oscillating with frequency $\om_0$ but with an amplitude-modified frequency. The magnitude of this correction was determined in the previous section. Equivalently, we  let the leading order approximation to be a wave that travels not with velocity $c_0$ but  $c_0+c_2$, with $c_2 \ll c_0$:
\beq
u^{(1)} (x,t) = -\frac{S_0}{2 \ttom_0} \cos (\om_0t - \ttom_0 x)
\eeq
where $\ttom_0$ differs from $\tom_0 = \om_0/c_0$. The amplitude of this wave is fixed by the boundary condition imposed at the origin, Eq. (\ref{discori}). However, this function is not a solution of (\ref{eq:orderone1}) but of 
\beq
\label{eq:orderone2}
\frac{1}{(c_0 + c_2)^2} \frac{\partial^2 u^{(1)}}{\partial t^2} -  \frac{\partial^2 u^{(1)} }{\partial x^2} = S_0 \delta (x) \sin \om_0 t \\
\eeq
with $\ttom_0 \equiv \om_0/(c_0 + c_2)$. Or, in Fourier space, 
\beq
\frac{1}{(c_0+c_2)^2} \frac{d^2}{dt^2}  u^{(1)}(k) + k^2   u^{(1)}(k) = S_0 \sin \om_0t
\label{loinhomc2}
\eeq
whose solution is a modification of (\ref{eq:uonek}):
\beq
 \hat u^{(1)}(k)  = \hat u^{(1)}_+(k) + {\hat u^{(1)}_-(k)}
 \label{eq:uonekc2}
\eeq
with
\beq
\hat u^{(1)}_{\pm}(k,t)= \pm \frac{ (S_0/2i) e^{\pm i\om_0 t +\epsilon t}}{ k^2 -(\tilde{\tom}_0 \mp i\epsilon)^2} 
\label{sol:orderonekc2}
\eeq
so that the privileged oscillators, the ones that are being driven at resonance, respond with an amplitude
\beq
\left.  \hat u^{(1)}(k,t) \right|_{k=\pm \tom_0} =  \frac{S_0}{4i}  \frac{1}{\tom_0^2 (c_2/c_0)} \left[ e^{ i\om_0t} - e^{- i\om_0t}  \right]
\eeq
which is very large compared to the oscillators with $k \ne \pm \tom_0$, since $(c_2/c_0)$ is very small. We identify these two oscillators with the ones that were considered in the previous section when computing the amplitude-dependent frequency:
\beq
\hat v^{(0)}(k,t)= {\cal A}  \left[ \delta (\tk -1) - \delta ( \tk +1) \right] \left[ e^{ic_0 kt} -  e^{-ic_0 kt} \right]
\eeq
with
\beq
{\cal A} \equiv \frac{S_0}{{4}i}  \frac{1}{\tom_0^2 (c_2/c_0)} \hspace{3em} \mbox{and \hspace{1em} $\tk \equiv k/k_0 = c_0k/\om_0$} \, .
\label{calavsc2}
\eeq
Comparing with (\ref{eq:homsol0}) we see that
\beq
A_k^{(0)} =  {\cal A}  \left[ \delta (\tk -1) - \delta ( \tk +1) \right]  = -B_k^{(0)}  \, ,
\eeq
so that, using \eqref{finalc2} and \eqref{calavsc2} we have 
\beq
\left(  \frac{c_2}{c_0} \right)^3 = - \frac{S_0^2}{16{\pi^2}} \left[  \beta^2 + \gamma \right] \, .
\label{eq:c2explicit}
\eeq

 With this information at hand we reformulate the successive approximation scheme including from the beginning the fact that, to leading order, the $k$-th oscillator does not oscillate with a frequency $c_0k$ but with a frequency $(c_0+c_2)k$: 
 \begin{small}
\bea
&&\hspace{-4em} \frac{1}{(c_0+c_2)^2} \frac{d^2}{dt^2} \left[ \hu^{(1)}(k) \right]  + \frac{1}{c^2}\frac{d^2}{dt^2} \left[ \hu^{(2)}(k) + \hu^{(3)}(k) + \dots \right] \nonumber \\
&&+ k^2 \left[  \hu^{(1)}(k) + \hu^{(2)}(k) + \hu^{(3)}(k) + \dots \right]
=  S_0 \sin \om_0t   \label{eq:wnspacec2}\\
&&{ \hspace{-4em} -\frac{i\beta k}{2\pi}  \int_{-\infty}^{\infty} dk' k' (k-k')  \left[ \hu^{(1)}(k') + \hu^{(2)}(k')  + \dots \right]   \left[ \hu^{(1)}(k-k') + \hu^{(2)}(k-k')  + \dots \right] \nonumber }  \\
&&  \hspace{-3em}  +\frac{\gamma k}{{(2\pi)^2}} \int_{-\infty}^{\infty} dk'  (k-k') [ \hat u^{(1)}(k-k')+\dots ]  \int_{-\infty}^{\infty} dk''  (k'-k'') k''  [\hat u^{(1)}(k'-k'')+\dots ][ \hat u^{(1)} (k'') +\dots ]  \nonumber
\eea
\end{small}
whose leading order solution has already been given in \eqref{eq:uonekc2}.
To go back to position space we take the FT of (\ref{eq:uonekc2}) and note that this is just (\ref{eq:sol_o1}) with the $\tom_0$ replaced by $\ttom_0$:
\bea
\label{eq:sol_o1corr}
\hu^{(1)}(x,t) &=& -  \frac{S_0}{2\ttom_0} \cos \left( \om_0 t-\ttom_0 |x| \right)     \\
&\approx &   -  \frac{S_0}{2\tom_0}  \left(1+ \frac{c_2}{c_0}   \right)   \cos \left( \om_0 t-\tom_0 (1-c_2/c_0) |x| \right)   
\eea
This is a monochromatic plane wave with speed of propagation $c_0(1+c_2/c_0)$. There is also a correction to the amplitude. Said amplitude, to leading order, is
\beq
\mA_1 \equiv \frac{S_0}{2\tom_0} { \left(1+ \frac{c_2}{c_0}   \right)}
\label{eq:leadingamp}
\eeq

\subsection{Second order}
\label{sec:2ndorderinhom}
If $u^{(1)}(k) $ is the solution to ({\ref{loinhomc2}),  the next order equation is
\begin{small}
\bea
\frac{d^2 \hat u^{(2)}(k)}{c_0^2d t^2} + k^2 \hat u^{(2)}(k) &=&  -\frac{i \beta k}{2\pi}  \int_{-\infty}^{\infty} dk' k' (k-k') \hat u^{(1)}(k') \hat u^{(1)}(k-k')      \\
&=&  -\frac{i \beta k}{2\pi}  \int_{-\infty}^{\infty} dk' k' (k-k') \left[  \hat u^{(1)}_+(k') + {\hat u^{(1)}_-(k')} \right] \left[  \hat u^{(1)}_+(k-k') + {\hat u^{(1)}_-(k-k')} \right] \nonumber
\eea
\end{small}
and we need{, using (\ref{sol:orderonekc2}) 
\begin{small}
\bea
\frac{d^2 \hat u_{++}^{(2)}(k)}{c_0^2d t^2} + k^2 \hat u_{++}^{(2)}(k) &=&  -\frac{i \beta k}{2\pi}  \int_{-\infty}^{\infty} dk' k' (k-k')   \hat u^{(1)}_+(k')  \hat u^{(1)}_+(k-k')   \\
&=& \frac{i\beta S_0^2k }{8\pi} e^{2i\om_0 t +2\epsilon t}   \int_{-\infty}^{\infty} dk' \frac{k'(k-k')}{[k'^2 -({\tilde{\tom}}_0 -i\epsilon)^2][(k-k')^2 -({\tilde{\tom}}_0 -i\epsilon)^2]}   \nonumber \\
& = & {+} \frac{\beta S_0^2 }{4}  \frac{k\tom_0 e^{2i\om_0 t +2\epsilon t}}{ k^2 -4(\tilde{\tom}_0 -i\epsilon)^2}
\label{eq:2nd++}
\eea
\end{small}
{where we have neglected a term of order $(c_2/c_0)$ in the numerator, that is, in the amplitude of the driver of the second harmonic}. {The} solution {to (\ref{eq:2nd++})}  is
\beq
\hat u_{++}^{(2)}(k) =  {+} \frac{\beta S_0^2 }{4} \frac{k\tom_0 e^{2i\om_0 t +2\epsilon t}}{[ k^2 -4({\tom}_0 -i\epsilon)^2] [ k^2 -4({\tilde{\tom}}_0 -i\epsilon)^2]}  \, .
\eeq

Similarly 
\bea
\frac{d^2 \hat u_{--}^{(2)}(k)}{c^2d t^2} + k^2 \hat u_{--}^{(2)}(k) &=&  -\frac{i \beta k}{2\pi}  \int_{-\infty}^{\infty} dk' k' (k-k')  {\hat u^{(1)}_-(k')} \,  {\hat u^{(1)}_-(k-k')}   
\eea
and
\beq
\hat u_{--}^{(2)}(k) =  {-} \frac{\beta S_0^2 }{4} \frac{k\tom_0 e^{-2i\om_0 t +2\epsilon t}}{[ k^2 -4({\tom}_0 +i\epsilon)^2] [ k^2 -4({\tilde{\tom}}_0 +i\epsilon)^2]}    
\eeq

We also need 
\bea
\label{pendiente}
\frac{d^2 \hat u_{+-}^{(2)}(k)}{c^2d t^2} + k^2 \hat u_{+-}^{(2)}(k) &=&  -\frac{i \beta k}{2\pi}  \int_{-\infty}^{\infty} dk' k' (k-k')   \hat u^{(1)}_+(k')  {\hat u^{(1)}_-(k-k')}    \\
&=& -\frac{i\beta S_0^2k }{8\pi}    \int_{-\infty}^{\infty} dk' \frac{k'(k-k')}{[k'^2 -({\tilde{\tom}}_0 -i\epsilon)^2][(k-k')^2 -({\tilde{\tom}}_0 +i\epsilon)^2]}   \nonumber  \\
&=& 0
\eea
and
\beq
 \hat u_{+-}^{(2)}(k) =  \hat u_{-+}^{(2)}(k) = 0.
\eeq
{so that
\beq
\label{u2hatk}
 \hat u^{(2)}(k) = \hat u_{++}^{(2)}(k) + \hat u_{--}^{(2)}(k) \, .
\eeq
}

Let us consider now waves traveling to the right for $x>0$. This indicates that the integral on the complex $k$-plane needed to obtain $u(x,t)$ has to be closed along the upper-half plane. This picks poles where the real part of $k$ is negative when $\om_0$ is positive:
\newpage
\bea
u_{++}^{(2)} (x,t) &=& \int_{-\infty}^{\infty} \frac{dk}{2\pi} \hat u_{++}^{(2)}(k) e^{ikx}    \\
&=& \frac{\beta S_0^2 \tom_0 }{4} \int_{-\infty}^{\infty} \frac{dk}{2\pi} \frac{k e^{ikx} e^{2i\om_0 t} }{[ k^2 -4({\tom}_0 -i\epsilon)^2] [ k^2 -4({\tilde{\tom}}_0 -i\epsilon)^2]}   \\
&=& \frac{i\beta S_0^2  }{{64}\, \tom_0 (c_2/c_0)}  \left[ e^{-2i \tom_0 x}  - e^{-2i {\tilde{\tom}}_0 x} \right] e^{2i\om_0 t} \\
&=& \frac{\beta S_0^2  }{{32}\, \tom_0 (c_2/c_0)}  e^{2i\om_0 t -2i \tom_0 (1-c_2/2c_0) x} \sin \tom_0 x c_2/c_0   \, .
\eea
The same reasoning leads to
\bea
u_{--}^{(2)} (x,t) 
&=&  \frac{\beta S_0^2  }{{32}\, \tom_0 (c_2/c_0)}  e^{-2i\om_0 t +2i \tom_0 (1-c_2/2c_0) x} \sin \tom_0 x c_2/c_0
\eea

Finally
\begin{align}
\label{eq:order2ohnezero}
\hspace{-2em}  {u^{(2)} (x,t) =}& \, u_{++}^{(2)}(x,t) + u_{--}^{(2)} (x,t)  \nonumber \\
=& \,  \frac{\beta S_0^2  }{{16}\, \tom_0 (c_2/c_0)} \cos \left[ 2\om_0 t -2 \tom_0 (1-c_2/2c_0) |x|  \right] \sin \left[ \tom_0 x c_2/c_0  \right]   \\
 \approx & \, \frac{\beta S_0^2  x}{{16}} \cos \left[ 2\om_0 t -2 \tom_0 (1-c_2/2c_0) |x|  \right]    \hspace{2em} \mbox{when $\frac{\tom_0 x c_2}{c_0} \ll 1$}
\label{eq:order2ohnezerosmaalx}
\end{align}

Consequently the second order solution is given by (\ref{eq:order2ohnezero}). It represents a plane wave of frequency $2\om_0$ traveling at speed $c_0+c_2/2$, whose amplitude is modulated by another wave of wavenumber $\tom_0 c_2/2c_0$. The amplitude of the latter is, using (\ref{eq:c2explicit}) and (\ref{eq:leadingamp}) 
\beq
\mA_2 \equiv \frac{\beta S_0^2}{{16}\, \tom_0 (c_2/c_0)} = \frac{S_0^{4/3}{\pi^{2/3}}}{2^{8/3}\, \tom_0} \frac{\beta}{ \left[    \beta^2  + \gamma \right]^{1/3} }  = \frac{  \tom_0^{1/3} {\pi^{2/3}} \beta}{{2}^{4/3} \left[   \beta^2  +\gamma\right]^{1/3} } \, \mA_1^{4/3}  \, .
\eeq
That is, the second order amplitude of the modulating wave is proportional to the leading order amplitude to the four third\flc{s}. The coefficient of proportionality involves both the second order ($\beta$) and the third order ($\gamma$) elastic constants. 

It is interesting to check what happens at short distances from the source, $x \tom_0 c_2 \ll c_0$. In this case, from (\ref{eq:order2ohnezerosmaalx}) we have 
\beq
\mA_{2 \, \, {\rm small} \, \, x} = \frac{\beta S_0^2 x}{{16}} = \frac{1}{{4}} \beta \, x \,  \tom_0^2 \,  \mA_1^2
\eeq
in agreement with the value quoted in the literature {(see for example Eq. (34) in Ref. \cite{McCall1994} or Eq. (34) in Ref. \cite{Wang2017})}. Note however that the notion of ``small $x$'' depends on the amplitude of the leading order wave, the frequency of the driving load, as well as on the nonlinear elastic constants:
\beq
x \ll \frac{(2\pi)^{2/3}}{\tom_0^{5/3}} \frac{1}{\left( \mA_1 \right)^{2/3}} \frac{1}{ \left[  \beta^2  +\gamma \right]^{1/3} } \, .
\label{eq:condsmallx}
\eeq
Figure \ref{Figure1} illustrates the main features of the second order solution.
\begin{figure*}[h]
{\includegraphics[width=\columnwidth]{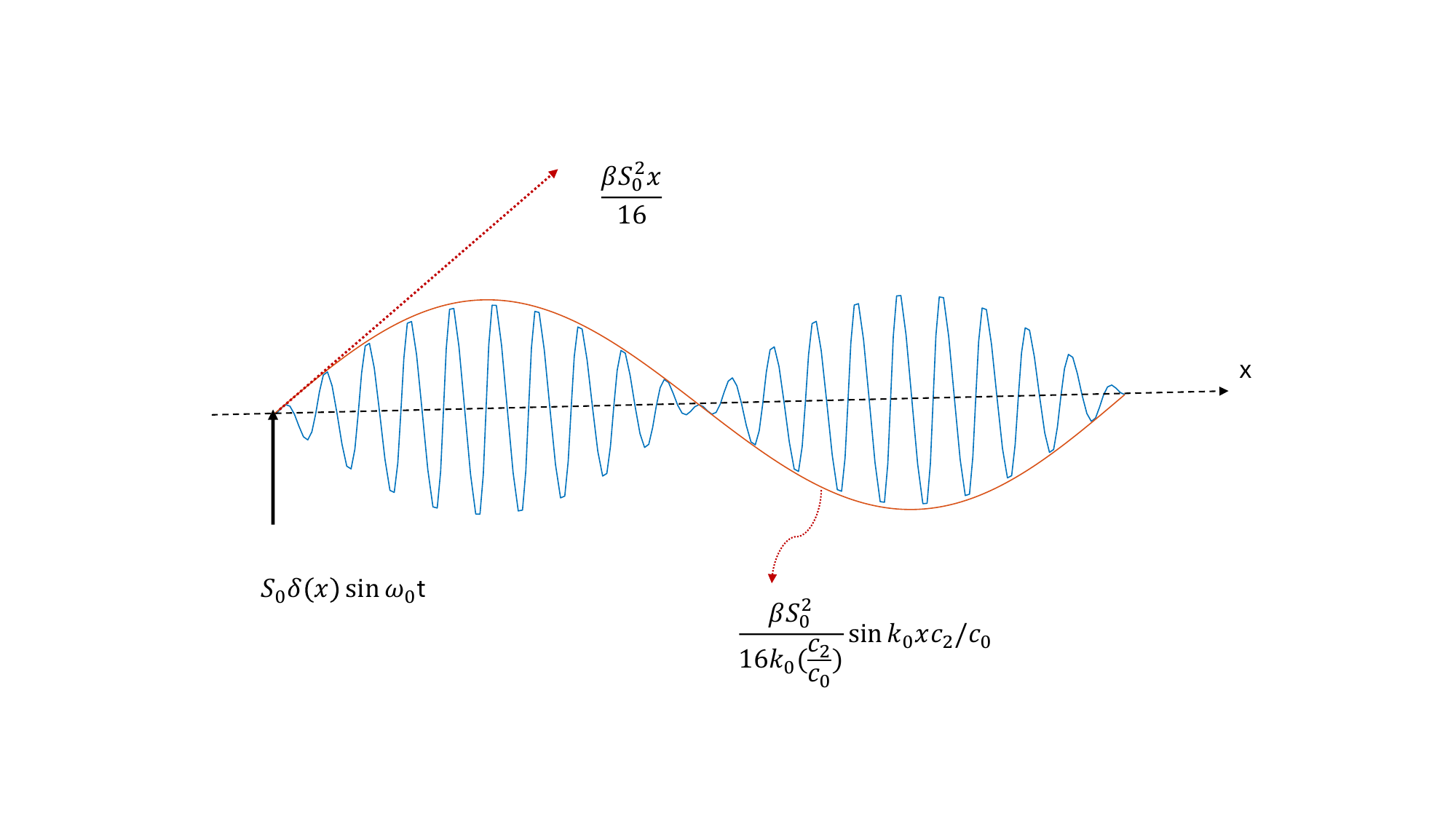}}
\caption{Schematics of the solution, Eq. \eqref{eq:order2ohnezero}, for the second-order approximation to the response of a one-dimensional nonlinear elastic medium loaded at the origin by a forcing periodic-in-time, Eq. \eqref{eq:fund}. It is a wave with twice the loading frequency (SHG), whose amplitude is modulated in distance (not in time). At short distaces from the souce, Eq. \eqref{eq:order2ohnezerosmaalx} is recovered. The modulation wavelength depends on the forcing amplitude, and on the material nonlinear constants, Eq. \eqref{eq:c2explicit}. The figure is a snapshot of Eq. \eqref{eq:order2ohnezero} taken at time $t=\pi/4\tom_0$}.
\label{Figure1}
\end{figure*}

To conclude this section, we note that the second harmonic of a sound wave carries information not only about the third order elastic constants, but also about the fourth order elastic constants. The latter will also appear in the generation of third order harmonics, to which we now turn our attention.

\subsection{Third order} 
One important motivation to study second harmonic generation is to measure second harmonic amplitudes and use the resulting measurement to infer the second order constant $\beta$ for a given material. As we have seen in the last section, the amplitude of the second harmonic by itself is not enough since it only provides a combination of $\beta$ and $\gamma$. To disentangle the two, it is necessary to measure the third harmonic as well. 

The equation to be solved is
\begin{small}
\bea
\frac{d^2 \hat u^{(3)}(k)}{c^2d t^2} + k^2   u^{(3)}(k) &=& -\frac{i\beta k}{2\pi}  \int_{-\infty}^{\infty} dk' k' (k-k')  \left[ \hu^{(1)}(k') + \hu^{(2)}(k')  \right]   \left[ \hu^{(1)}(k-k') + \hu^{(2)}(k-k')  \right] \nonumber  \\
&&  \hspace{-3em}  + \frac{\gamma k}{{(2\pi)^2}} \int_{-\infty}^{\infty} dk'  (k-k') [ \hat u^{(1)}(k-k') ]  \int_{-\infty}^{\infty} dk''  (k'-k'') k''  [\hat u^{(1)}(k'-k'') ][ \hat u^{(1)} (k'')  ] \nonumber  \\
&& 
\label{eqorder3}
\eea
\end{small}
with $\hu^{(1)}(k) $ given by (\ref{eq:sol_o1corr}) and $\hu^{(2)}(k')$ by (\ref{eq:order2ohnezerosmaalx}) with $c_2/c_0$ given by (\ref{eq:c2explicit}). A rather tedious but straightforward calculation using repeatedly Cauchy's theorem leads to 
\beq
u^{(3)} (x,t)  =   u^{(3)}_{\beta 3} (x,t) + u^{(3)}_{\beta 1} (x,t) + u^{(3)}_{\gamma 3} (x,t) + u^{(3)}_{\gamma 1} (x,t)
\eeq
where 
\bea
\label{beta3}
 u^{(3)}_{\beta 3} (x,t) &=&  \frac{\beta^2 S_0^3}{{64}\, \tom_0 (c_2/c_0)^2} \left[ \frac 23 \cos 3(\om_0 t - \tom_0 |x|) - \cos 3(\om_0t -\tom_0[1-c_2/3c_0]|x|) \right. \nonumber  \\
&& \hspace{4em} +\left. \frac 13 \cos 3(\om_0 t - \tom_0 [1-c_2/c_0]|x|)  \right]  \\
& \approx & -\frac{\beta^2 S_0^3 \tom_0 x^2}{{64}} \cos 3(\om_0 t - \tom_0 |x|)    \hspace{6em} \mbox{when $\frac{\tom_0 x c_2}{c_0} \ll 1$}   \\
\label{beta1}
 u^{(3)}_{\beta 1} (x,t)  &=& \frac{\beta^2 S_0^3}{{16}\, \tom_0  (c_2/c_0)^2}  \cos (\om_0 t -\tom_0 |x|)  \sin^2 \tom_0 x c_2/2c_0   \\
& \approx & \frac{\beta^2 S_0^3 \tom_0 x^2}{{64}} \cos (\om_0 t -\tom_0 |x|) \hspace{7em} \mbox{when $\frac{\tom_0 x c_2}{c_0} \ll 1$}  \\
\label{gamma3}
u^{(3)}_{\gamma 3} (x,t) & = &{-} \frac{ \gamma S_0^3}{{96} \, \tom_0 (c_2/c_0) }  \sin \left[ 3\om_0 t -3 \tom_0 (1-c_2/2c_0) |x|  \right]  \sin 3\tom_0 |x| c_2/2c_0  \\
& \approx & {-} \frac{  \gamma S_0^3 x}{{64} }  \sin \left[ 3\om_0 t -3 \tom_0 (1-c_2/2c_0) |x|  \right]     \hspace{2.5em} \mbox{when $\frac{\tom_0 x c_2}{c_0} \ll 1$}   \\
\label{gamma1}
u^{(3)}_{\gamma 1} (x,t) & = &  \frac{ \gamma S_0^3}{{32} \, \tom_0 (c_2/c_0) }  \sin \left[ \om_0 t - \tom_0 (1-c_2/2c_0) |x|  \right]  \sin \tom_0 |x| c_2/2c_0  \\
& \approx &  \frac{ \gamma S_0^3 |x|}{{64} }  \sin \left[ \om_0 t - \tom_0 (1-c_2/2c_0) |x|  \right]     \hspace{4em} \mbox{when $\frac{\tom_0 |x| c_2}{c_0} \ll 1$}
\eea

Thus, the third order solution to Eq. (\ref{eq:fund}) is the sum of four terms. Two of them are third harmonics, evolving in time as $3\om_0t$, and the other two evolve as $\om_0t$. In both cases one of them is proportional to $\beta^2$ and the other to $\gamma$. The ``$\beta$'' terms scale with the amplitude of the source as $S_0^{5/3}$ and the ``$\gamma$'' terms scale as $S_0^{7/3}$. 

\section{Discussion and outlook}
We have provided a solution to the problem of an acoustic wave propagating along one dimension in a weakly nonlinear elastic medium, loaded at the origin by a monochromatic periodic forcing of frequency $\om_0$ and amplitude $S_0$, Eq. (\ref{eq:fund}), by successive approximations when the amplitude of the forcing is small. The leading order solution, Eq. (\ref{eq:sol_o1corr}), is a plane wave whose speed of propagation is modified from the linear $c_0$ to $c_0+c_2$, with $c_2/c_0$ given by (\ref{eq:c2explicit}), a modification that depends on the amplitude of the loading and on the nonlinear parameters of the medium. The second order solution is given by (\ref{eq:order2ohnezero}). It is a plane wave of frequency $2\om_0$, a second harmonic, whose amplitude depends both on the second order $\beta$ and third order $\gamma$ coefficients, and is modulated by a wave of long wavelength, proportional to $c_0/c_2$. The amplitude of this second harmonic scales with the amplitude $\mA_1$ of the leading term as $\mA_1^{4/3}$ and not as its amplitude squared, as is the case for free nonlinear oscillations.    At short distances from the origin, the expression often found in the literature is recovered. The third order solution is the sum of four terms, Eqs. (\ref{beta3}), (\ref{beta1}), (\ref{gamma3}) and (\ref{gamma1}). They are amplitude-modulated plane waves, with long modulation wavelengths, proportional to $c_0/c_2$. Two of them are of frequency $3\om_0$, third harmonics, and the other two are of frequency $\omega_0$.

The amplitude of the second harmonic solution involves not only the distance from the source and the second order parameter $\beta$, but the third order parameter $\gamma$ as well. This means that measuring the amplitude of the second harmonic as a function of distance from the source can be used to determine both $\beta$ and $\gamma$.

A natural next step to take, using the reasoning reported herein, could be to study the behavior of the response of a weakly nonlinear elastic medium to a loading that is transverse rather than longitudinal. As indicated, for example by Wang and Achenbah \cite{Wang2017}, the nonlinear coupling between both modes could lead to interesting insights of use for non-destructive testing. After that, nonlinear Rayleigh waves, also of importance in nondestructive testing, present themselves as attractive candidates for study. And after that, nonlinear waves in plates and rods.

\section*{Acknowledgements}
Numerous stimulating discussions  with Nicol\'as Mujica, Vicente Salinas, Carolina Espinoza and Javiera Gonz\'alez are gratefully acknowledged, as is the collaboration of David  Esp\'\i ndola in the early stages of this project, as well as instructive interactions with Claudio Falc\'on, Rodrigo Vicencio and Bruno Scheihing-Hitschfeld.  This work was supported by Fondecyt Grants 1191179, 1230938, and U. de Chile VID Grant ENL12/22.

%\section*{References}
%apsrev4-2.bst 2019-01-14 (MD) hand-edited version of apsrev4-1.bst
%Control: key (0)
%Control: author (8) initials jnrlst
%Control: editor formatted (1) identically to author
%Control: production of article title (0) allowed
%Control: page (0) single
%Control: year (1) truncated
%Control: production of eprint (0) enabled
%

%\bibliography{NonlinSHG}

\begin{thebibliography}{29}%
\makeatletter
\providecommand \@ifxundefined [1]{%
 \@ifx{#1\undefined}
}%
\providecommand \@ifnum [1]{%
 \ifnum #1\expandafter \@firstoftwo
 \else \expandafter \@secondoftwo
 \fi
}%
\providecommand \@ifx [1]{%
 \ifx #1\expandafter \@firstoftwo
 \else \expandafter \@secondoftwo
 \fi
}%
\providecommand \natexlab [1]{#1}%
\providecommand \enquote  [1]{``#1''}%
\providecommand \bibnamefont  [1]{#1}%
\providecommand \bibfnamefont [1]{#1}%
\providecommand \citenamefont [1]{#1}%
\providecommand \href@noop [0]{\@secondoftwo}%
\providecommand \href [0]{\begingroup \@sanitize@url \@href}%
\providecommand \@href[1]{\@@startlink{#1}\@@href}%
\providecommand \@@href[1]{\endgroup#1\@@endlink}%
\providecommand \@sanitize@url [0]{\catcode `\\12\catcode `\$12\catcode
  `\&12\catcode `\#12\catcode `\^12\catcode `\_12\catcode `\%12\relax}%
\providecommand \@@startlink[1]{}%
\providecommand \@@endlink[0]{}%
\providecommand \url  [0]{\begingroup\@sanitize@url \@url }%
\providecommand \@url [1]{\endgroup\@href {#1}{\urlprefix }}%
\providecommand \urlprefix  [0]{URL }%
\providecommand \Eprint [0]{\href }%
\providecommand \doibase [0]{https://doi.org/}%
\providecommand \selectlanguage [0]{\@gobble}%
\providecommand \bibinfo  [0]{\@secondoftwo}%
\providecommand \bibfield  [0]{\@secondoftwo}%
\providecommand \translation [1]{[#1]}%
\providecommand \BibitemOpen [0]{}%
\providecommand \bibitemStop [0]{}%
\providecommand \bibitemNoStop [0]{.\EOS\space}%
\providecommand \EOS [0]{\spacefactor3000\relax}%
\providecommand \BibitemShut  [1]{\csname bibitem#1\endcsname}%
\let\auto@bib@innerbib\@empty
%</preamble>
\bibitem [{\citenamefont {Matlack}\ \emph {et~al.}(2014)\citenamefont
  {Matlack}, \citenamefont {Kim}, \citenamefont {Jacobs},\ and\ \citenamefont
  {Qu}}]{Matlack2014}%
  \BibitemOpen
  \bibfield  {author} {\bibinfo {author} {\bibfnamefont {K.~H.}\ \bibnamefont
  {Matlack}}, \bibinfo {author} {\bibfnamefont {J.~Y.}\ \bibnamefont {Kim}},
  \bibinfo {author} {\bibfnamefont {L.~J.}\ \bibnamefont {Jacobs}},\ and\
  \bibinfo {author} {\bibfnamefont {J.}~\bibnamefont {Qu}},\ }\bibfield
  {title} {\bibinfo {title} {Review of second harmonic generation measurement
  techniques for material state determination in metals},\ }\href
  {https://doi.org/10.1007/s10921-014-0273-5} {\bibfield  {journal} {\bibinfo
  {journal} {Journal of Nondestructive Evaluation}\ }\textbf {\bibinfo {volume}
  {34}},\ \bibinfo {pages} {273} (\bibinfo {year} {2014})}\BibitemShut
  {NoStop}%
\bibitem [{\citenamefont {Achenbach}\ and\ \citenamefont
  {Wang}(2018)}]{Achenbach2018}%
  \BibitemOpen
  \bibfield  {author} {\bibinfo {author} {\bibfnamefont {J.~D.}\ \bibnamefont
  {Achenbach}}\ and\ \bibinfo {author} {\bibfnamefont {Y.}~\bibnamefont
  {Wang}},\ }\bibfield  {title} {\bibinfo {title} {Far-field resonant third
  harmonic surface wave on a half-space of incompressible material of cubic
  nonlinearity},\ }\href
  {https://doi.org/https://doi.org/10.1016/j.jmps.2017.09.010} {\bibfield
  {journal} {\bibinfo  {journal} {Journal of the Mechanics and Physics of
  Solids}\ }\textbf {\bibinfo {volume} {120}},\ \bibinfo {pages} {5} (\bibinfo
  {year} {2018})},\ \bibinfo {note} {special issue in honor of Ares J. Rosakis
  on the occasion of his 60th birthday}\BibitemShut {NoStop}%
\bibitem [{\citenamefont {Sahu}\ \emph {et~al.}(2017)\citenamefont {Sahu},
  \citenamefont {Swaminathan}, \citenamefont {Bandhoypadhyay},\ and\
  \citenamefont {Sagar}}]{Sahu2017}%
  \BibitemOpen
  \bibfield  {author} {\bibinfo {author} {\bibfnamefont {M.~K.}\ \bibnamefont
  {Sahu}}, \bibinfo {author} {\bibfnamefont {J.}~\bibnamefont {Swaminathan}},
  \bibinfo {author} {\bibfnamefont {N.}~\bibnamefont {Bandhoypadhyay}},\ and\
  \bibinfo {author} {\bibfnamefont {S.}~\bibnamefont {Sagar}},\ }\bibfield
  {title} {\bibinfo {title} {Rayleigh surface wave based non linear ultrasound
  to assess effect of precipitation hardening during tempering in p92 steel},\
  }\href {https://doi.org/https://doi.org/10.1016/j.msea.2017.07.014}
  {\bibfield  {journal} {\bibinfo  {journal} {Materials Science and
  Engineering: A}\ }\textbf {\bibinfo {volume} {703}},\ \bibinfo {pages} {76}
  (\bibinfo {year} {2017})}\BibitemShut {NoStop}%
\bibitem [{\citenamefont {Fuchs}\ \emph {et~al.}(2021)\citenamefont {Fuchs},
  \citenamefont {Qu}, \citenamefont {Kim}, \citenamefont {Unocic},
  \citenamefont {Guo}, \citenamefont {Ramuhalli},\ and\ \citenamefont
  {Jacobs}}]{Fuchs2021}%
  \BibitemOpen
  \bibfield  {author} {\bibinfo {author} {\bibfnamefont {B.}~\bibnamefont
  {Fuchs}}, \bibinfo {author} {\bibfnamefont {J.}~\bibnamefont {Qu}}, \bibinfo
  {author} {\bibfnamefont {J.-Y.}\ \bibnamefont {Kim}}, \bibinfo {author}
  {\bibfnamefont {K.~A.}\ \bibnamefont {Unocic}}, \bibinfo {author}
  {\bibfnamefont {Q.}~\bibnamefont {Guo}}, \bibinfo {author} {\bibfnamefont
  {P.}~\bibnamefont {Ramuhalli}},\ and\ \bibinfo {author} {\bibfnamefont
  {L.~J.}\ \bibnamefont {Jacobs}},\ }\bibfield  {title} {\bibinfo {title}
  {Analytical modeling of the evolution of the nonlinearity parameter of
  sensitized stainless steel},\ }\href {https://doi.org/10.1063/5.0053632}
  {\bibfield  {journal} {\bibinfo  {journal} {Journal of Applied Physics}\
  }\textbf {\bibinfo {volume} {130}},\ \bibinfo {pages} {165102} (\bibinfo
  {year} {2021})}\BibitemShut {NoStop}%
\bibitem [{\citenamefont {Bellotti}\ \emph {et~al.}(2021)\citenamefont
  {Bellotti}, \citenamefont {Kim}, \citenamefont {Bishop}, \citenamefont
  {Jared}, \citenamefont {Johnson}, \citenamefont {Susan}, \citenamefont
  {Noell},\ and\ \citenamefont {Jacobs}}]{Belotti2021}%
  \BibitemOpen
  \bibfield  {author} {\bibinfo {author} {\bibfnamefont {A.}~\bibnamefont
  {Bellotti}}, \bibinfo {author} {\bibfnamefont {J.-Y.}\ \bibnamefont {Kim}},
  \bibinfo {author} {\bibfnamefont {J.~E.}\ \bibnamefont {Bishop}}, \bibinfo
  {author} {\bibfnamefont {B.~H.}\ \bibnamefont {Jared}}, \bibinfo {author}
  {\bibfnamefont {K.}~\bibnamefont {Johnson}}, \bibinfo {author} {\bibfnamefont
  {D.}~\bibnamefont {Susan}}, \bibinfo {author} {\bibfnamefont {P.~J.}\
  \bibnamefont {Noell}},\ and\ \bibinfo {author} {\bibfnamefont {L.~J.}\
  \bibnamefont {Jacobs}},\ }\bibfield  {title} {\bibinfo {title} {Nonlinear
  ultrasonic technique for the characterization of microstructure in additive
  materials},\ }\href {https://doi.org/10.1121/10.0002960} {\bibfield
  {journal} {\bibinfo  {journal} {The Journal of the Acoustical Society of
  America}\ }\textbf {\bibinfo {volume} {149}},\ \bibinfo {pages} {158}
  (\bibinfo {year} {2021})}\BibitemShut {NoStop}%
\bibitem [{\citenamefont {Espinoza}\ \emph {et~al.}(2018)\citenamefont
  {Espinoza}, \citenamefont {Feli{\'u}}, \citenamefont {Aguilar}, \citenamefont
  {Espinoza-Gonz{\'a}ez}, \citenamefont {Lund}, \citenamefont {Salinas},\ and\
  \citenamefont {Mujica}}]{Espinoza2018}%
  \BibitemOpen
  \bibfield  {author} {\bibinfo {author} {\bibfnamefont {C.}~\bibnamefont
  {Espinoza}}, \bibinfo {author} {\bibfnamefont {D.}~\bibnamefont {Feli{\'u}}},
  \bibinfo {author} {\bibfnamefont {C.}~\bibnamefont {Aguilar}}, \bibinfo
  {author} {\bibfnamefont {R.}~\bibnamefont {Espinoza-Gonz{\'a}ez}}, \bibinfo
  {author} {\bibfnamefont {F.}~\bibnamefont {Lund}}, \bibinfo {author}
  {\bibfnamefont {V.}~\bibnamefont {Salinas}},\ and\ \bibinfo {author}
  {\bibfnamefont {N.}~\bibnamefont {Mujica}},\ }\bibfield  {title} {\bibinfo
  {title} {Linear versus nonlinear acoustic probing of plasticity in metals: A
  quantitative assessment},\ }\bibfield  {journal} {\bibinfo  {journal}
  {Materials}\ }\textbf {\bibinfo {volume} {11}},\ \href
  {https://doi.org/10.3390/ma11112217} {10.3390/ma11112217} (\bibinfo {year}
  {2018})\BibitemShut {NoStop}%
\bibitem [{\citenamefont {Sosa}\ \emph {et~al.}(2024)\citenamefont {Sosa},
  \citenamefont {Carvajal}, \citenamefont {Salinas~Barrera}, \citenamefont
  {Lund}, \citenamefont {Aguilar},\ and\ \citenamefont
  {Castro~Cerda}}]{Sosa2024}%
  \BibitemOpen
  \bibfield  {author} {\bibinfo {author} {\bibfnamefont {M.}~\bibnamefont
  {Sosa}}, \bibinfo {author} {\bibfnamefont {L.}~\bibnamefont {Carvajal}},
  \bibinfo {author} {\bibfnamefont {V.}~\bibnamefont {Salinas~Barrera}},
  \bibinfo {author} {\bibfnamefont {F.}~\bibnamefont {Lund}}, \bibinfo {author}
  {\bibfnamefont {C.}~\bibnamefont {Aguilar}},\ and\ \bibinfo {author}
  {\bibfnamefont {F.}~\bibnamefont {Castro~Cerda}},\ }\bibfield  {title}
  {\bibinfo {title} {Acoustic assessment of microstructural deformation
  mechanisms on a cold rolled {C}u30{Z}n brass},\ }\href@noop {} {\bibfield
  {journal} {\bibinfo  {journal} {Materials}\ }\textbf {\bibinfo {volume}
  {17}},\ \bibinfo {pages} {3321} (\bibinfo {year} {2024})}\BibitemShut
  {NoStop}%
\bibitem [{\citenamefont {Chakrapani}\ and\ \citenamefont
  {Barnard}(2017)}]{Chakrapani2017}%
  \BibitemOpen
  \bibfield  {author} {\bibinfo {author} {\bibfnamefont {S.~K.}\ \bibnamefont
  {Chakrapani}}\ and\ \bibinfo {author} {\bibfnamefont {D.~J.}\ \bibnamefont
  {Barnard}},\ }\bibfield  {title} {\bibinfo {title} {Determination of acoustic
  nonlinearity parameter ({$\beta$}) using nonlinear resonance ultrasound
  spectroscopy: Theory and experiment},\ }\href
  {https://doi.org/10.1121/1.4976057} {\bibfield  {journal} {\bibinfo
  {journal} {The Journal of the Acoustical Society of America}\ }\textbf
  {\bibinfo {volume} {141}},\ \bibinfo {pages} {919} (\bibinfo {year}
  {2017})}\BibitemShut {NoStop}%
\bibitem [{\citenamefont {Meo}\ \emph {et~al.}(2008)\citenamefont {Meo},
  \citenamefont {Polimeno},\ and\ \citenamefont {Zumpano}}]{Meo2008}%
  \BibitemOpen
  \bibfield  {author} {\bibinfo {author} {\bibfnamefont {M.}~\bibnamefont
  {Meo}}, \bibinfo {author} {\bibfnamefont {U.}~\bibnamefont {Polimeno}},\ and\
  \bibinfo {author} {\bibfnamefont {G.}~\bibnamefont {Zumpano}},\ }\bibfield
  {title} {\bibinfo {title} {Detecting damage in composite material using
  nonlinear elastic wave spectroscopy methods},\ }\href@noop {} {\bibfield
  {journal} {\bibinfo  {journal} {Applied Composite Materials}\ }\textbf
  {\bibinfo {volume} {15}},\ \bibinfo {pages} {115} (\bibinfo {year}
  {2008})}\BibitemShut {NoStop}%
\bibitem [{\citenamefont {Muller}\ \emph {et~al.}(2005)\citenamefont {Muller},
  \citenamefont {Sutin}, \citenamefont {Guyer}, \citenamefont {Talmant},
  \citenamefont {Laugier},\ and\ \citenamefont {Johnson}}]{Muller2005}%
  \BibitemOpen
  \bibfield  {author} {\bibinfo {author} {\bibfnamefont {M.}~\bibnamefont
  {Muller}}, \bibinfo {author} {\bibfnamefont {A.}~\bibnamefont {Sutin}},
  \bibinfo {author} {\bibfnamefont {R.}~\bibnamefont {Guyer}}, \bibinfo
  {author} {\bibfnamefont {M.}~\bibnamefont {Talmant}}, \bibinfo {author}
  {\bibfnamefont {P.}~\bibnamefont {Laugier}},\ and\ \bibinfo {author}
  {\bibfnamefont {P.~A.}\ \bibnamefont {Johnson}},\ }\bibfield  {title}
  {\bibinfo {title} {Nonlinear resonant ultrasound spectroscopy (nrus) applied
  to damage assessment in bone},\ }\href@noop {} {\bibfield  {journal}
  {\bibinfo  {journal} {The Journal of the Acoustical Society of America}\
  }\textbf {\bibinfo {volume} {118}},\ \bibinfo {pages} {3946} (\bibinfo {year}
  {2005})}\BibitemShut {NoStop}%
\bibitem [{\citenamefont {Haupert}\ \emph {et~al.}(2015)\citenamefont
  {Haupert}, \citenamefont {Gu{\'e}rard}, \citenamefont {Mitton}, \citenamefont
  {Peyrin},\ and\ \citenamefont {Laugier}}]{Haupert2015}%
  \BibitemOpen
  \bibfield  {author} {\bibinfo {author} {\bibfnamefont {S.}~\bibnamefont
  {Haupert}}, \bibinfo {author} {\bibfnamefont {S.}~\bibnamefont
  {Gu{\'e}rard}}, \bibinfo {author} {\bibfnamefont {D.}~\bibnamefont {Mitton}},
  \bibinfo {author} {\bibfnamefont {F.}~\bibnamefont {Peyrin}},\ and\ \bibinfo
  {author} {\bibfnamefont {P.}~\bibnamefont {Laugier}},\ }\bibfield  {title}
  {\bibinfo {title} {Quantification of nonlinear elasticity for the evaluation
  of submillimeter crack length in cortical bone},\ }\href@noop {} {\bibfield
  {journal} {\bibinfo  {journal} {Journal of the mechanical behavior of
  biomedical materials}\ }\textbf {\bibinfo {volume} {48}},\ \bibinfo {pages}
  {210} (\bibinfo {year} {2015})}\BibitemShut {NoStop}%
\bibitem [{\citenamefont {Payan}\ \emph {et~al.}(2007)\citenamefont {Payan},
  \citenamefont {Garnier}, \citenamefont {Moysan},\ and\ \citenamefont
  {Johnson}}]{Payan2007}%
  \BibitemOpen
  \bibfield  {author} {\bibinfo {author} {\bibfnamefont {C.}~\bibnamefont
  {Payan}}, \bibinfo {author} {\bibfnamefont {V.}~\bibnamefont {Garnier}},
  \bibinfo {author} {\bibfnamefont {J.}~\bibnamefont {Moysan}},\ and\ \bibinfo
  {author} {\bibfnamefont {P.}~\bibnamefont {Johnson}},\ }\bibfield  {title}
  {\bibinfo {title} {Applying nonlinear resonant ultrasound spectroscopy to
  improving thermal damage assessment in concrete},\ }\href@noop {} {\bibfield
  {journal} {\bibinfo  {journal} {The Journal of the Acoustical Society of
  America}\ }\textbf {\bibinfo {volume} {121}},\ \bibinfo {pages} {EL125}
  (\bibinfo {year} {2007})}\BibitemShut {NoStop}%
\bibitem [{\citenamefont {Payan}\ \emph {et~al.}(2014)\citenamefont {Payan},
  \citenamefont {Ulrich}, \citenamefont {Le~Bas}, \citenamefont {Saleh},\ and\
  \citenamefont {Guimaraes}}]{Payan2014}%
  \BibitemOpen
  \bibfield  {author} {\bibinfo {author} {\bibfnamefont {C.}~\bibnamefont
  {Payan}}, \bibinfo {author} {\bibfnamefont {T.~J.}\ \bibnamefont {Ulrich}},
  \bibinfo {author} {\bibfnamefont {P.-Y.}\ \bibnamefont {Le~Bas}}, \bibinfo
  {author} {\bibfnamefont {T.}~\bibnamefont {Saleh}},\ and\ \bibinfo {author}
  {\bibfnamefont {M.}~\bibnamefont {Guimaraes}},\ }\bibfield  {title} {\bibinfo
  {title} {Quantitative linear and nonlinear resonance inspection techniques
  and analysis for material characterization: Application to concrete thermal
  damage},\ }\href@noop {} {\bibfield  {journal} {\bibinfo  {journal} {The
  Journal of the Acoustical Society of America}\ }\textbf {\bibinfo {volume}
  {136}},\ \bibinfo {pages} {537} (\bibinfo {year} {2014})}\BibitemShut
  {NoStop}%
\bibitem [{\citenamefont {Persson}\ \emph {et~al.}(2020)\citenamefont
  {Persson}, \citenamefont {Haller}, \citenamefont {Karlsson},\ and\
  \citenamefont {Koz{\l}owski}}]{Persson2020}%
  \BibitemOpen
  \bibfield  {author} {\bibinfo {author} {\bibfnamefont {K.}~\bibnamefont
  {Persson}}, \bibinfo {author} {\bibfnamefont {K.}~\bibnamefont {Haller}},
  \bibinfo {author} {\bibfnamefont {S.}~\bibnamefont {Karlsson}},\ and\
  \bibinfo {author} {\bibfnamefont {M.}~\bibnamefont {Koz{\l}owski}},\
  }\bibfield  {title} {\bibinfo {title} {Non-destructive testing of the
  strength of glass by a non-linear ultrasonic method},\ }in\ \href@noop {}
  {\emph {\bibinfo {booktitle} {Challenging Glass Conference Proceedings}}},\
  Vol.~\bibinfo {volume} {7}\ (\bibinfo {year} {2020})\BibitemShut {NoStop}%
\bibitem [{\citenamefont {Kube}\ and\ \citenamefont
  {Arguelles}(2017)}]{Kube2017}%
  \BibitemOpen
  \bibfield  {author} {\bibinfo {author} {\bibfnamefont {C.~M.}\ \bibnamefont
  {Kube}}\ and\ \bibinfo {author} {\bibfnamefont {A.~P.}\ \bibnamefont
  {Arguelles}},\ }\bibfield  {title} {\bibinfo {title} {Ultrasonic harmonic
  generation from materials with up to cubic nonlinearity},\ }\href@noop {}
  {\bibfield  {journal} {\bibinfo  {journal} {The Journal of the Acoustical
  Society of America}\ }\textbf {\bibinfo {volume} {142}},\ \bibinfo {pages}
  {EL224} (\bibinfo {year} {2017})}\BibitemShut {NoStop}%
\bibitem [{\citenamefont {Lissenden}(2021)}]{Lissenden2021}%
  \BibitemOpen
  \bibfield  {author} {\bibinfo {author} {\bibfnamefont {C.~J.}\ \bibnamefont
  {Lissenden}},\ }\bibfield  {title} {\bibinfo {title} {Nonlinear ultrasonic
  guided waves---principles for nondestructive evaluation},\ }\href@noop {}
  {\bibfield  {journal} {\bibinfo  {journal} {Journal of Applied Physics}\
  }\textbf {\bibinfo {volume} {129}} (\bibinfo {year} {2021})}\BibitemShut
  {NoStop}%
\bibitem [{\citenamefont {Jhang}\ \emph {et~al.}(2020)\citenamefont {Jhang},
  \citenamefont {Choi},\ and\ \citenamefont {Kim}}]{Jhang2020}%
  \BibitemOpen
  \bibfield  {author} {\bibinfo {author} {\bibfnamefont {K.-Y.}\ \bibnamefont
  {Jhang}}, \bibinfo {author} {\bibfnamefont {S.}~\bibnamefont {Choi}},\ and\
  \bibinfo {author} {\bibfnamefont {J.}~\bibnamefont {Kim}},\ }\bibfield
  {title} {\bibinfo {title} {Measurement of nonlinear ultrasonic parameters
  from higher harmonics},\ }in\ \href@noop {} {\emph {\bibinfo {booktitle}
  {Measurement of Nonlinear Ultrasonic Characteristics}}},\ \bibinfo {editor}
  {edited by\ \bibinfo {editor} {\bibfnamefont {K.-Y.}\ \bibnamefont {Jhang}},
  \bibinfo {editor} {\bibfnamefont {C.~J.}\ \bibnamefont {Lissenden}}, \bibinfo
  {editor} {\bibfnamefont {I.}~\bibnamefont {Solodov}}, \bibinfo {editor}
  {\bibfnamefont {Y.}~\bibnamefont {Ohara}},\ and\ \bibinfo {editor}
  {\bibfnamefont {V.}~\bibnamefont {Gusev}}}\ (\bibinfo  {publisher}
  {Springer},\ \bibinfo {year} {2020})\ pp.\ \bibinfo {pages}
  {9--60}\BibitemShut {NoStop}%
\bibitem [{\citenamefont {Park}\ \emph {et~al.}(2021)\citenamefont {Park},
  \citenamefont {Kim}, \citenamefont {Song}, \citenamefont {Choi},\ and\
  \citenamefont {Jhang}}]{Park2021}%
  \BibitemOpen
  \bibfield  {author} {\bibinfo {author} {\bibfnamefont {S.-H.}\ \bibnamefont
  {Park}}, \bibinfo {author} {\bibfnamefont {J.}~\bibnamefont {Kim}}, \bibinfo
  {author} {\bibfnamefont {D.-G.}\ \bibnamefont {Song}}, \bibinfo {author}
  {\bibfnamefont {S.}~\bibnamefont {Choi}},\ and\ \bibinfo {author}
  {\bibfnamefont {K.-Y.}\ \bibnamefont {Jhang}},\ }\bibfield  {title} {\bibinfo
  {title} {Measurement of absolute acoustic nonlinearity parameter using
  laser-ultrasonic detection},\ }\href@noop {} {\bibfield  {journal} {\bibinfo
  {journal} {Applied Sciences}\ }\textbf {\bibinfo {volume} {11}},\ \bibinfo
  {pages} {4175} (\bibinfo {year} {2021})}\BibitemShut {NoStop}%
\bibitem [{\citenamefont {Ostrovsky}\ and\ \citenamefont
  {Johnson}(2001)}]{Ostrovsky2001}%
  \BibitemOpen
  \bibfield  {author} {\bibinfo {author} {\bibfnamefont {L.~A.}\ \bibnamefont
  {Ostrovsky}}\ and\ \bibinfo {author} {\bibfnamefont {P.~A.}\ \bibnamefont
  {Johnson}},\ }\bibfield  {title} {\bibinfo {title} {Dynamic nonlinear
  elasticity in geomaterials},\ }\href {https://doi.org/10.1007/BF03548898}
  {\bibfield  {journal} {\bibinfo  {journal} {La Rivista del Nuovo Cimento}\
  }\textbf {\bibinfo {volume} {24}},\ \bibinfo {pages} {1} (\bibinfo {year}
  {2001})}\BibitemShut {NoStop}%
\bibitem [{\citenamefont {McCall}(1994)}]{McCall1994}%
  \BibitemOpen
  \bibfield  {author} {\bibinfo {author} {\bibfnamefont {K.~R.}\ \bibnamefont
  {McCall}},\ }\bibfield  {title} {\bibinfo {title} {Theoretical study of
  nonlinear elastic wave propagation},\ }\href
  {https://doi.org/https://doi.org/10.1029/93JB02974} {\bibfield  {journal}
  {\bibinfo  {journal} {Journal of Geophysical Research: Solid Earth}\ }\textbf
  {\bibinfo {volume} {99}},\ \bibinfo {pages} {2591} (\bibinfo {year}
  {1994})}\BibitemShut {NoStop}%
\bibitem [{\citenamefont {Martin}\ \emph {et~al.}(2018)\citenamefont {Martin},
  \citenamefont {Bodet}, \citenamefont {Tournat},\ and\ \citenamefont
  {Rejiba}}]{Martin2019}%
  \BibitemOpen
  \bibfield  {author} {\bibinfo {author} {\bibfnamefont {R.}~\bibnamefont
  {Martin}}, \bibinfo {author} {\bibfnamefont {L.}~\bibnamefont {Bodet}},
  \bibinfo {author} {\bibfnamefont {V.}~\bibnamefont {Tournat}},\ and\ \bibinfo
  {author} {\bibfnamefont {F.}~\bibnamefont {Rejiba}},\ }\bibfield  {title}
  {\bibinfo {title} {{Seismic wave propagation in nonlinear viscoelastic media
  using the auxiliary differential equation method}},\ }\href
  {https://doi.org/10.1093/gji/ggy441} {\bibfield  {journal} {\bibinfo
  {journal} {Geophysical Journal International}\ }\textbf {\bibinfo {volume}
  {216}},\ \bibinfo {pages} {453} (\bibinfo {year} {2018})}\BibitemShut
  {NoStop}%
\bibitem [{\citenamefont {O'Brien}(2020)}]{OBrien2020}%
  \BibitemOpen
  \bibfield  {author} {\bibinfo {author} {\bibfnamefont {G.~S.}\ \bibnamefont
  {O'Brien}},\ }\bibfield  {title} {\bibinfo {title} {{A lattice method for
  seismic wave propagation in nonlinear viscoelastic media}},\ }\href
  {https://doi.org/10.1093/gji/ggaa537} {\bibfield  {journal} {\bibinfo
  {journal} {Geophysical Journal International}\ }\textbf {\bibinfo {volume}
  {224}},\ \bibinfo {pages} {1572} (\bibinfo {year} {2020})}\BibitemShut
  {NoStop}%
\bibitem [{\citenamefont {De~Lima}\ and\ \citenamefont
  {Hamilton}(2003)}]{deLima2003}%
  \BibitemOpen
  \bibfield  {author} {\bibinfo {author} {\bibfnamefont {W.}~\bibnamefont
  {De~Lima}}\ and\ \bibinfo {author} {\bibfnamefont {M.}~\bibnamefont
  {Hamilton}},\ }\bibfield  {title} {\bibinfo {title} {Finite-amplitude waves
  in isotropic elastic plates},\ }\href@noop {} {\bibfield  {journal} {\bibinfo
   {journal} {Journal of sound and vibration}\ }\textbf {\bibinfo {volume}
  {265}},\ \bibinfo {pages} {819} (\bibinfo {year} {2003})}\BibitemShut
  {NoStop}%
\bibitem [{\citenamefont {Nayfeh}\ and\ \citenamefont
  {Mook}(2008)}]{Nayfeh2008}%
  \BibitemOpen
  \bibfield  {author} {\bibinfo {author} {\bibfnamefont {A.~H.}\ \bibnamefont
  {Nayfeh}}\ and\ \bibinfo {author} {\bibfnamefont {D.~T.}\ \bibnamefont
  {Mook}},\ }\href@noop {} {\emph {\bibinfo {title} {Nonlinear oscillations}}}\
  (\bibinfo  {publisher} {John Wiley \& Sons},\ \bibinfo {year}
  {2008})\BibitemShut {NoStop}%
\bibitem [{\citenamefont {Swinteck}\ \emph {et~al.}(2013)\citenamefont
  {Swinteck}, \citenamefont {Muralidharan},\ and\ \citenamefont
  {Deymier}}]{Swinteck2013}%
  \BibitemOpen
  \bibfield  {author} {\bibinfo {author} {\bibfnamefont {N.~Z.}\ \bibnamefont
  {Swinteck}}, \bibinfo {author} {\bibfnamefont {K.}~\bibnamefont
  {Muralidharan}},\ and\ \bibinfo {author} {\bibfnamefont {P.~A.}\ \bibnamefont
  {Deymier}},\ }\bibfield  {title} {\bibinfo {title} {Phonon scattering in
  one-dimensional anharmonic crystals and superlattices: Analytical and
  numerical study},\ }\href@noop {} {\bibfield  {journal} {\bibinfo  {journal}
  {Journal of vibration and acoustics}\ }\textbf {\bibinfo {volume} {135}},\
  \bibinfo {pages} {041016} (\bibinfo {year} {2013})}\BibitemShut {NoStop}%
\bibitem [{\citenamefont {Manktelow}\ \emph {et~al.}(2011)\citenamefont
  {Manktelow}, \citenamefont {Leamy},\ and\ \citenamefont
  {Ruzzene}}]{Manktelow2011}%
  \BibitemOpen
  \bibfield  {author} {\bibinfo {author} {\bibfnamefont {K.}~\bibnamefont
  {Manktelow}}, \bibinfo {author} {\bibfnamefont {M.~J.}\ \bibnamefont
  {Leamy}},\ and\ \bibinfo {author} {\bibfnamefont {M.}~\bibnamefont
  {Ruzzene}},\ }\bibfield  {title} {\bibinfo {title} {Multiple scales analysis
  of wave--wave interactions in a cubically nonlinear monoatomic chain},\
  }\href@noop {} {\bibfield  {journal} {\bibinfo  {journal} {Nonlinear
  Dynamics}\ }\textbf {\bibinfo {volume} {63}},\ \bibinfo {pages} {193}
  (\bibinfo {year} {2011})}\BibitemShut {NoStop}%
\bibitem [{\citenamefont {Zarembo}\ and\ \citenamefont
  {Krasil'nikov}(1971)}]{Zarembo1971}%
  \BibitemOpen
  \bibfield  {author} {\bibinfo {author} {\bibfnamefont {L.~K.}\ \bibnamefont
  {Zarembo}}\ and\ \bibinfo {author} {\bibfnamefont {V.~A.}\ \bibnamefont
  {Krasil'nikov}},\ }\bibfield  {title} {\bibinfo {title} {Nonlinear phenomena
  in the propagation of elastic waves in solids},\ }\href@noop {} {\bibfield
  {journal} {\bibinfo  {journal} {Soviet Physics Uspekhi}\ }\textbf {\bibinfo
  {volume} {13}},\ \bibinfo {pages} {778} (\bibinfo {year} {1971})}\BibitemShut
  {NoStop}%
\bibitem [{\citenamefont {Khoo}\ and\ \citenamefont {Wang}(1976)}]{Khoo1976}%
  \BibitemOpen
  \bibfield  {author} {\bibinfo {author} {\bibfnamefont {I.~C.}\ \bibnamefont
  {Khoo}}\ and\ \bibinfo {author} {\bibfnamefont {Y.~K.}\ \bibnamefont
  {Wang}},\ }\bibfield  {title} {\bibinfo {title} {Multiple time scale analysis
  of an anharmonic crystal},\ }\href {https://doi.org/10.1063/1.522884}
  {\bibfield  {journal} {\bibinfo  {journal} {Journal of Mathematical Physics}\
  }\textbf {\bibinfo {volume} {17}},\ \bibinfo {pages} {222} (\bibinfo {year}
  {1976})}\BibitemShut {NoStop}%
\bibitem [{\citenamefont {Wang}\ and\ \citenamefont
  {Achenbach}(2017)}]{Wang2017}%
  \BibitemOpen
  \bibfield  {author} {\bibinfo {author} {\bibfnamefont {Y.}~\bibnamefont
  {Wang}}\ and\ \bibinfo {author} {\bibfnamefont {J.~D.}\ \bibnamefont
  {Achenbach}},\ }\bibfield  {title} {\bibinfo {title} {Interesting effects in
  harmonic generation by plane elastic waves},\ }\href@noop {} {\bibfield
  {journal} {\bibinfo  {journal} {Acta Mechanica Sinica}\ }\textbf {\bibinfo
  {volume} {33}},\ \bibinfo {pages} {754} (\bibinfo {year} {2017})}\BibitemShut
  {NoStop}%
\end{thebibliography}

\end{document}